\documentclass{article}

\usepackage[margin=1.1in]{geometry}
\usepackage{authblk}
\usepackage{graphicx}
\usepackage{subfigure}
\usepackage{amsmath,bm}
\usepackage{gensymb}
\usepackage{amsfonts}
\usepackage{advdate}

\begin{document}

\title{\huge{Thermodynamics and computation\\during collective motion near criticality}}

\author[]{Emanuele Crosato}
\author[]{Richard E. Spinney}
\author[]{Ramil Nigmatullin}
\author[]{Joseph T. Lizier}
\author[]{Mikhail Prokopenko}
\affil[]{Complex Systems Research Group and Centre for Complex Systems, Faculty of Engineering \& IT, The University of Sydney, Sydney, NSW 2006, Australia.}

\date{\AdvanceDate[-3]\today}

\maketitle

\begin{abstract}
We study self-organisation of collective motion as a thermodynamic phenomenon, in the context of the first law of thermodynamics.
It is expected that the coherent ordered motion typically self-organises in the presence of changes in the (generalised) internal energy and of (generalised) work done on, or extracted from, the system.
We aim to explicitly quantify changes in these two quantities in a system of simulated self-propelled particles, and contrast them with changes in the system's configuration entropy.
In doing so, we adapt a thermodynamic formulation of the curvatures of the internal energy and the work, with respect to two parameters that control the particles' alignment.
This allows us to systematically investigate the behaviour of the system by varying the two control parameters to drive the system across a kinetic phase transition.
Our results identify critical regimes and show that during the phase transition, where the configuration entropy of the system decreases, the rates of change of the work and of the internal energy also decrease, while their curvatures diverge.
Importantly, the reduction of entropy achieved through expenditure of work is shown to peak at criticality.
We relate this both to a thermodynamic efficiency and the significance of the increased order with respect to a computational path.
Additionally, this study provides an information-geometric interpretation of the curvature of the internal energy as the difference between two curvatures: the curvature of the free entropy, captured by the Fisher information, and the curvature of the configuration entropy.
\end{abstract}


\section{Introduction}


Collective motion involves self-organisation of coherent movement in a system of self-propelled particles, and is a pervasive phenomenon observed in many biological, chemical and physical systems~\cite{vicsek2012collective}.
Collective motion has been studied in animals (e.g., flocks of birds~\cite{ballerini2008interaction}, schools of fish~\cite{parrish2002self} and colonies of insects~\cite{buhl2006from}), in bacteria~\cite{sokolov2009enhanced}, in tissue cells~\cite{szabo2006phase}, in moving biomolecules~\cite{schaller2010polar} and even in non-living systems such as autonomous micromotors~\cite{ibele2009schooling}.
Despite their diversity, these systems can exhibit similar motion patterns, such as orientated aggregations, stationary clusters and swirls~\cite{vicsek2012collective}.
A crucial characteristic that distinguishes collective motion from other kinds of coordinated motion, is that complex patterns can self-organise from simple local interactions among individual particles, without requiring any global control or leading roles~\cite{camazine2001self}, but involving information cascades~\cite{couzin2007collective, wang2012quantifying}.
Nevertheless, systems of self-propelled particles can display remarkable dynamic coordination during collective motion, as well as other interesting features, such as scalability,
response to the environment and reconfiguration after external intrusions.


The ubiquity of collective motion, and its similarity across different systems, suggest the existence of underlying universal principles, the investigation of which has become a well-established, cross-disciplinary pursuit.
The formulation of general laws bridging local interactions and group-level properties is one of the main challenges for defining a unified theory of collective motion~\cite{couzin2009collective}.


A first step towards this goal was the conception of dynamical models~\cite{vicsek1995novel, toner1995long, toner1998flocks, gregoire2004onset}.
Vicsek et al.~\cite{vicsek1995novel} introduced a dynamical model of collective motion inspired by ferromagnetism, in which particles assume the average direction of motion of other particles in its neighbourhood (similarly to magnetisation), with some random perturbation (similarly to temperature).
The authors simulated the motion for gradually decreasing random perturbation and observed a kinetic phase transition between a 
disorderly moving phase and a phase with coherent (oriented) motion, the critical point of which was localised using a suitable order parameter.
Several studies have followed Vicsek's intuition, and extensions of the model have been proposed.
Gr\'{e}goire and Chat\'{e}~\cite{gregoire2004onset}, for example, studied the effect of several control parameters on the collective behaviour of a modified version of Vicsek's model, which adds a cohesion component to the motion rules.
The authors confirmed the existence of the kinetic phase transition and, by varying the strength of the additional cohesion component, observed three more phases: a ``gas'', a ``liquid'' and a ``solid'' phase, also separated by phase transitions.


More recently, Bialek et al.~\cite{bialek2012statistical, bialek2014social, castellana2016entropic} provided a statistical mechanical model for the propagation of directional order throughout flocks.
On the hypothesis that flocks have statistically stationary states, the authors calculated the maximum entropy distribution~\cite{jaynes1957information} of birds' normalised velocities, consistent with the average pairwise directional correlation experimentally observed from the field data (i.e., large flocks of \textit{Sturnus vulgarishas}~\cite{ballerini2008empirical, cavagna2008starflag, cavagna2008starflag2}).
Bialek's statistical mechanical description provides a formal theoretical framework to make quantitative predictions of emergent collective phenomena.
For instance, the model was shown to be capable of predicting the existence of pairwise correlations on all length scales, as well as four-body correlations~\cite{bialek2012statistical}.
The model was also shown to be capable of predicting the flight directions of birds in the interior of the flock, given the directions of the birds on the border.


Despite this fundamental contribution, current statistical mechanical approaches to collective motion do not explicitly incorporate thermodynamic quantities such as free entropy and work, dynamics of which are especially important during phase transitions.
In this article we aim to investigate this quantities in the dynamical model of collective motion proposed by Gr\'{e}goire and Chat\'{e}~\cite{gregoire2004onset}, which undergoes a kinetic phase transition over parameters that control the particles' alignment: from a ``disordered motion'' phase, in which particles keep changing direction but occupy a fairly stable collective space, to a ``coherent motion'' phase, in which particles cohesively move towards a common direction.
The control parameters that we consider are the alignment strength among particles and the number of nearest neighbours affecting a particle's alignment.
A quasi-static process is considered, during which these two control parameters are varied infinitesimally slowly, driving the system across the phase transition while thermodynamic equilibrium is maintained.
The dynamics of fundamental thermodynamical quantities, such as the \emph{generalised} work, heat and energy, are investigated over the quasi-static process, in the context of the first law of thermodynamics.


The choice of a quasi-static protocol allows the application of our theoretical framework, which requires the system to be in a steady state.
Moreover, the results obtained considering a quasi-static protocol can be meaningfully interpreted in the context of the second law of thermodynamics, to get useful insights into more realistic processes.
For instance, the work done on the system in the quasi-static limit is a lower bound for the work that would be done on the system using a protocol in which the control parameter is varied faster.


In this study, we use a method that allows us to give a statistical mechanical interpretation of the curvatures of the generalised work and of the generalised internal energy of the system with respect to the control parameter.
Such method exploits the relationship between these two curvatures and two information-theoretic quantities, the configuration entropy and the Fisher information (a measure of the information that an observed variable carries about the parameter), which can be numerically estimated by simulating the system using different values of the control parameters.


We also provide two information-geometric expressions of the curvature of the internal energy and related quantities with respect to the control parameter.
On the one hand, the curvature of the internal energy is proportional to the \emph{difference} between two curvatures: the curvature of the free entropy, captured by the Fisher information, and the curvature of the configuration entropy.
This expression highlights a ``computational balance'' present in distributed computational processes, of which collective motion is an example.
Such balance relates the \emph{sensitivity} of the system to changes in control parameter (captured by the Fisher information) and the system's \emph{uncertainty} (captured by the configuration entropy).
This enhances the view of the ``thermodynamic balance'', reflected by the first law in the context of quasi-static processes, between the configuration entropy of the system, its internal energy and the work done on, or extracted from, the system.
On the other hand, we derive another quantity as the \emph{sum} of the Fisher information and the curvature of the configuration entropy.


Our computational results show that, in the simulated system of particles during collective motion, the rates of change of the generalised work and of the generalised internal energy decrease with the control parameters, whenever the system of self-propelled particles begins to move more coherently.
This dynamic is particularly steep near criticality, where the curvatures of these quantities with respect to the control parameters are shown to diverge.
The configuration entropy of the system is shown to decrease during the phase transition, as the system self-organises into a more ordered phase.
The thermodynamic perspective adopted in this study allows us to define a notion of thermodynamic efficiency of computation as a ratio of entropy changes to the required work.
In addition, we propose an interpretation of this work rate as a distance along a computational path implied by the control parameter, measured in terms of the cumulative sensitivity to the changes in the control parameter.
Specifically, our results suggest that the reduction of the configuration entropy, indicating the increase in the internal order within the considered collective motion, is most significant at criticality.


In addition to these main results, this paper confirms and quantifies critical dynamics in statistical mechanical models of collective motion, which were previously observed in dynamical models~\cite{vicsek1995novel, gregoire2004onset}.
Moreover, it is shown that the Fisher information diverges at criticality, and can therefore be used to build a phase diagram of the dynamics of the system.


The remainder of this article is structured as follows.
Section~\ref{sec:background} provides the technical preliminaries necessary for understanding the role of the Fisher information in physical systems, the information-geometrical interpretation of the studied curvatures, the quasi-static protocol that we consider and the dynamical model of collective motion.
Section~\ref{sec:results} presents our statistical mechanical formulation of the curvatures of the generalised work and internal energy, and the computational results of simulated collective motion.
The results are discussed in Section~\ref{sec:conclusions}, where our conclusions are also provided.


\section{Technical preliminaries}
\label{sec:background}


\subsection{Fisher information and physical systems}
\label{sec:fisher}


The Fisher information~\cite{fisher1922mathematical} is a known quantity in statistics and information theory.
It measures the amount of information that an observable random variable $X$ carries about an unknown parameter $\theta$.
For many parameters $\theta = [\theta_1,\theta_2,\dots,\theta_M]^T$, the Fisher information matrix is defined as
\begin{equation}
\label{eq:fisher-matrix}
F_{mn}(\theta) = E \Bigg[ \bigg( \frac{\partial \ln p(x|\theta)}{\partial\theta_m} \bigg) \bigg( \frac{\partial \ln p(x|\theta)}{\partial\theta_n} \bigg) \Bigg| \theta \Bigg] ,
\end{equation}
where $p(x|\theta)$ is the probability of the realisation $x$ of $X$ given the parameters $\theta$, and the function $E(y)$ is the expected value of $y$.


In recent years, the meaning of the Fisher information for physical systems has been investigated in thermodynamical and statistical mechanical terms~\cite{brody1995geometrical, brody2003information, janke2004information, crooks2007measuring, crooks2011fisher, prokopenko2011relating, machta2013parameter, prokopenko2015information}.
Let us consider a physical system, described by the state functions $X_m(x)$ over the configuration space.
The probability of the states of the system, in a stationary state, is given by the Gibbs measure:
\begin{equation}
\label{eq:gibbs-measure}
p(x|\theta) = \frac{1}{Z(\theta)}e^{-\beta H(x,\theta)} = \frac{1}{Z(\theta)}e^{-\sum_m \theta_m X_m(x)} ,
\end{equation}
where $\beta=1/k_bT$ is the inverse temperature $T$ ($k_b$ being the Boltzmann constant), the Hamiltonian $H(x,\theta)$ defines the total energy at state $x$, $\theta_m$ are thermodynamic variables (pressure, magnetic field, chemical potential, etc.) and $Z(\theta)$ is the partition function~\cite{brody1995geometrical, crooks2007measuring}.
The Gibbs free energy of such system is:
\begin{equation}
\label{eq:gibbs-potential}
G(T,\theta_m) = U(S,\phi_m) - TS - \phi_m\theta_m ,
\end{equation}
where $U$ is the internal energy of the system, $S$ is the configuration entropy and $\phi_m$ is an order parameter.
For a physical system described by the Gibbs measure in Equation \eqref{eq:gibbs-measure}, the Fisher information measures the size of the fluctuations about equilibrium in the collective variables $X_m$ and $X_n$ and is related to the derivatives of the corresponding order parameters with respect to the collective variables~\cite{crooks2011fisher,prokopenko2011relating}:
\begin{equation}
\begin{aligned}
 F_{mn}(\theta) &= \Big\langle (X_m(x) - \langle X_m \rangle ) (X_n(x) - \langle X_n \rangle ) \Big\rangle \\
 &= \beta\frac{\partial\phi_m}{\partial\theta_n} ,
 \end{aligned}
\end{equation}
where the angle brackets represent average values over the ensemble.


Fisher information has been related to entropy production~\cite{plastino1997relationship}, and also used as a variational principle to derive fundamental thermodynamic laws~\cite{frieden1992fisher, frieden1999fisher} or for predicting modelling~\cite{machta2013parameter}.


\subsection{Interpretation of entropic curvatures}
\label{sec:curvatures}


It has also been shown~\cite{brody1995geometrical, brody2003information, janke2004information, crooks2007measuring} that the Fisher information is equivalent to the thermodynamic metric tensor:
\begin{equation}
\label{eq:tensor}
 F_{mn}(\theta) = g_{mn}(\theta) = \frac{\partial^2\psi}{\partial\theta_m\partial\theta_n} ,
\end{equation}
where $\psi = \ln Z = -\beta G$ is the free entropy (for isothermal systems, $\psi$ is proportional to the free energy).
In other words, the Fisher information is the curvature of the free entropy $(\ln Z)$.
This reveals the information-geometrical meaning of the Fisher information as a Riemannian metric (more precisely, the Fisher-Rao metric) for the manifold of thermodynamic states, providing a measure of distance between thermodynamic states.
Thus, information-geometrically, the Fisher information can be interpreted as an average uncertainty density on a statistical manifold, proportional to the volume of geodesic balls~\cite{petz2002covariance}.


This study provides thermodynamical interpretations for curvatures, focussing specifically on quantities that can be computed numerically from the probability distribution of the observed variables, such as the configuration entropy $S$ of the system.
In particular, we propose the curvatures
\begin{equation}
\label{eq:curvatures-of-interest}
\frac{d^2(\mathbb{S})^{\pm}}{d\theta^2} \equiv \frac{d^2 S}{d\theta^2} \pm F(\theta)
\end{equation}
as the central quantities of interest (notice that a single control parameter $\theta$ is now used).
Therefore, the quantity $d^2(\mathbb{S})^{\pm} / d\theta^2$ is either the sum of, or the difference between, average statistical uncertainties (i.e., the volumes of geodesic balls) attributed to the free entropy and to the configuration entropy.


In order to interpret these information-geometric, static, quantities in terms of traditional thermodynamic quantities (e.g., heat and work, defined dynamically) we must give meaning to the notion of a change with respect to the control parameter, $\theta$, i.e. we must define the process or \emph{protocol}.
By protocol we mean a defined evolution of the control parameter in time, i.e. $\theta(t)$, which drives the system between different states and in doing so incurs changes in heat, work etc.
By establishing such a protocol we can give physical meaning to integrals of the curvatures $(\mathbb{S})^{\pm}$, such that ${d(\mathbb{S})^{\pm}}/{d\theta}$ can be readily interpreted as a change in $(\mathbb{S})^{\pm}$ under the action of the protocol. 
It is of critical importance to recognise that the nature of the protocol determines the physical behaviour of the quantity $(\mathbb{S})^{\pm}$, i.e., its decomposition into heat and work.
The most natural example is a quasi-static protocol, which we discuss next, though note that less conventional alternatives can be designed (as will be discussed in Section~\ref{sec:conclusions}).


\subsection{Quasi-static protocols}
\label{sec:quasi-static}


A quasi-static protocol is an idealised driving process carried out over an infinite amount of time, such that we can consider the system to be in equilibrium throughout the process.
For instance, a linear quasi-static protocol taking the system from a distribution characterised by $\theta^1$ to $\theta^2$, would be given by the limit
\begin{equation}
\theta(t) = \lim_{\tau\to\infty}\theta^1 + \frac{t}{\tau}(\theta^2 - \theta^1) .
\end{equation}
Since the system is always in equilibrium, the total entropy production of the universe (the system and the environment) is zero, and therefore any change in the configuration entropy due to the driving process is identically matched by a flow of heat that manifests as entropy change in the environment:
\begin{equation}
\label{eq:entropy-heat-qs}
\frac{dS}{d\theta} = \frac{d \langle\beta Q_{gen}\rangle}{d\theta} ,
\end{equation}
where a sign convention dictates that $Q_{gen}$ is the generalised heat flow \emph{from} the environment \emph{to} the system.
Here the subscript indicates a generalised heat in the sense of Jaynes~\cite{jaynes1957information}, such that we can consider
\begin{equation}
\label{eq:generalised}
\langle U_{gen}\rangle = U(S,\phi) - \phi\theta
\end{equation}
and the generalised first law holds $\Delta\langle U_{gen}\rangle = \Delta\langle Q_{gen}\rangle + \Delta\langle W_{gen}\rangle$, where $W_{gen}$ is the generalised work.
Equation~\eqref{eq:entropy-heat-qs} leads to a formulation of the first law of thermodynamics, in case of a quasi-static processes, as
\begin{equation}
\label{eq:first-law-quasi}
\frac{d \langle\beta U_{gen}\rangle}{d\theta} = \frac{d S}{d\theta} + \frac{d \langle\beta W_{gen}\rangle}{d\theta} .
\end{equation}


It is worth noting that, according to the second law of thermodynamics, a change in the free energy of the system requires a greater or equal amount of work to be done on the system, which is $\Delta \langle W_{gen}\rangle \ge \Delta G$.
In the quasi-static limit the work required is exactly the change in the free energy, therefore $\Delta\langle W_{gen}\rangle = \Delta G$.
In other words, the total work performed on the system (which can be calculated by integrating the infinitesimal work changes over a range of the control parameter) is a lower bound for the work that would be performed on the system if we were not considering the quasi-static limit.


This methodology is very general, provided a Gibbs form can be postulated and a probability distribution can be estimated.
However, in this study we focus on a system of self-propelled particles during collective motion, driven across a phase transition by a quasi-static protocol acting on two parameters that control the particles' alignment.
The model of collective motion that we adopted is presented in the following section.


\subsection{Dynamical and statistical mechanical models of collective motion}


We consider the model of collective motion proposed by Gr\'{e}goire and Chat\'{e}~\cite{gregoire2004onset}.
Let's have $N$ self-propelled particles.
At time $t$, each particle $i=\{1,2,\dots,N\}$ has position $\mathbf{x}_i(t)$ and velocity $\mathbf{v}_i(t)$.
The time evolution of position and velocity is given by the following rules:
\begin{equation}
\label{eq:cm-model-position}
\mathbf{x}_i(t+1) = \mathbf{x}_i(t) + \mathbf{v}_i(t) ,
\end{equation}
\begin{equation}
\label{eq:cm-model-velocity}
\mathbf{v}_i(t+1) = v_0\Theta \left[ a\sum_{j\in n_c^i}\mathbf{v}_j(t) + b\sum_{j\in n_c^i} f_{ij} + n_c\bm{\eta}_i \right] .
\end{equation}
The normalisation operator $\Theta(\mathbf{y})=\mathbf{y}/|\mathbf{y}|$ keeps the particles' speed constant, i.e., $|\mathbf{v}_i(t)|=v_0$ at every time $t$.
The argument of the normalisation operator is the sum of three velocity components: from left to right, we have an \emph{alignment}, a \emph{cohesion} and a \emph{perturbation} components.
The alignment component for particle $i$ is the sum of the velocities of its nearest neighbourhood $j \in n_c^i$ of fixed size $n_c$ (i.e., $n_c^i$ includes the $n_c$ particles with the smallest Euclidean distance from $i$, and is updated at each time step).
The cohesion component is the sum of the cohesion forces $f_{ij}$ between particle $i$ and its neighbours.
The parameters $a$ and $b$ are, respectively, the weights of the alignment and the cohesion components.
The perturbation is introduced by means of a random unit vector $\bm{\eta}_i$, and is weighted by the fixed number of nearest neighbours $n_c$ of each particle.


The forces $f_{ij}$ are functions of the distances $r_{ij}$:
\begin{equation}
\label{eq:cm-model-distance}
\begin{aligned}
f_{ij}(r_{ij} < r_b) &= -\infty\cdot\mathbf{e}_{ij} ,\\
f_{ij}(r_b \le r_{ij} < r_a) &= \frac{1}{4} \cdot \frac{r_{ij} - r_e}{r_a - r_e}\mathbf{e}_{ij} ,\\
f_{ij}(r_a \le r_{ij} < r_0) &= \mathbf{e}_{ij} ,
\end{aligned}
\end{equation}
where $r_b$, $r_e$, $r_a$ and $r_0$ are distance parameters (with $r_b < r_e < r_a < r_0$) and $\mathbf{e}_{ij}$ is the unit vector in the direction from $x_i(t)$ to $x_j(t)$, at time $t$.
When the distance $r_{ij}$ between two particles is within a ``repulsion'' limit $r_b$, particle $i$ moves away from particle $j$, towards the opposite direction of $\mathbf{e}_{ij}$.
When $r_{ij}$ is between the limits $r_a$ and $r_b$, particle $i$ adjusts its velocity in order to maintain an intermediate ``equilibrium'' distance $r_e$ from $j$ ($r_e$ is typically the average between $r_a$ and $r_b$).
When the distance $r_{ij}$ is larger than $r_a$, but smaller than $r_0$, particle $i$ modifies its velocity in order to get closer to $j$.
If particle $i$ is farther than $r_0$ from $j$, then $j$ does not affect the cohesion component of the velocity of $i$.


Collective motion can also be modelled using statistical mechanics, for example, by providing a formulation for the probability distribution of the velocities $\mathbf{v}_i$.
Bialek et. al~\cite{bialek2012statistical} defined a statistical mechanics model of collective motion that can describe flocking phenomena, including the dynamics in the model by Gr\'{e}goire and Chat\'{e}~\cite{gregoire2004onset}.
In its more general version, which does not take into consideration whether the particles are in the inner or outer region of the group, the statistical mechanical model is the following:
\begin{equation}
\label{eq:cm-model-bialek}
p(\mathbf{v}_i|J,n_c) = \frac{1}{Z(J,n_c)}\exp \left[ \frac{J}{2}\sum_{i=1}^{N}\sum_{j\in n_c^i}\mathbf{v}_i\cdot\mathbf{v}_j \right] ,
\end{equation}
where $Z$ is the partition function and $J = v_0a/n_c$ represents the alignment strength between particles.
Crucially, such model has plausible dynamics that allows the system to relax towards, and fluctuate around, an equilibrium, which is analogous to many dynamical models: particles move according to a weighted sum of neighbours' direction while being affected by a random perturbation.


\section{Method and results}
\label{sec:results}


\subsection{Relating information-theoretic and thermodynamic quantities in the quasi-static limit}
\label{sec:fisher-and-thermo}


Based on the relations presented in Sections~\ref{sec:fisher},~\ref{sec:curvatures} and~\ref{sec:quasi-static}, we use Equations~\eqref{eq:gibbs-potential},~\eqref{eq:tensor} and~\eqref{eq:generalised} to obtain
\begin{equation}
\label{eq:fisher-as-difference}
F(\theta) = - \frac{d^2\langle\beta U_{gen}\rangle}{d\theta^2} + \frac{d^2 S}{d\theta^2} ,
\end{equation}
which then leads to the definition of
\begin{equation}
\label{eq:fisher-quasi-static}
\frac{d^2(\mathbb{S})^-}{d\theta^2} \equiv \frac{d^2 S}{d\theta^2} - F(\theta) = \frac{d^2\langle\beta U_{gen}\rangle}{d\theta^2} .
\end{equation}
This expression, which is a key result of our study, makes it evident that the second derivative of the internal energy scaled by $\beta$ (expressed on the right-hand side) is proportional to the difference between two curvatures: the curvature of the free entropy, captured by the Fisher information, and the curvature of the configuration entropy.
It is important to note that Equation~\eqref{eq:fisher-as-difference} holds in general, since $U_{gen}$ is related, only, to the stationary distribution given by the Gibbs measure. 


However, the decomposition of $U_{gen}$ into $Q_{gen}$ and $W_{gen}$ ($\beta=1$) depends on the protocol.
Here we explicitly relate the Fisher information and the generalised work, energy and heat in systems driven by quasi-static protocols.
In the quasi-static limit, we show how the Fisher information can be related to the second derivative of the generalised work.
By further differentiating the first law for quasi-static processes in Equation~\eqref{eq:first-law-quasi} over the control parameter, and by expressing it for the work term, we obtain
\begin{equation}
\frac{d^2 \langle\beta W_{gen}\rangle}{d\theta^2} = \frac{d^2\langle\beta U_{gen}\rangle}{d\theta^2} - \frac{d^2 S}{d\theta^2} ,
\end{equation}
which, by comparison with Equation~\eqref{eq:fisher-as-difference}, leads to another important result:
\begin{equation}
\label{eq:fisher-work-curvature}
F(\theta) = - \frac{d^2\langle\beta W_{gen}\rangle}{d\theta^2} .
\end{equation}
In Equation~\eqref{eq:fisher-work-curvature}, the Fisher information has an information-geometric meaning at given values of $\theta$, while we have no physical interpretation for $\langle W_{gen}(\theta)\rangle$ unless we also specify a protocol and a path $\theta^0\to\theta$.
If we assume that $\theta$ increases, and thus $\theta > \theta^0$, we have
\begin{equation}
\label{eq:integration-fisher}
\begin{aligned}
\int_{\theta^0}^\theta F(\theta^\prime)d\theta^\prime &= -\int_{\theta^0}^\theta \frac{d^2\langle\beta W_{gen}\rangle}{d(\theta^\prime)^2}d\theta^\prime\\
&= -\frac{d\langle\beta W_{gen}\rangle}{d\theta} + c(\theta^0) .
\end{aligned}
\end{equation}
The value of $c(\theta^0)$ can be determined by identifying the value of the control parameter $\theta^*$, for which small changes incur no work, i.e.,
\begin{equation}
\label{eq:theta-star}
\left.\frac{d\langle\beta W_{gen}\rangle}{d\theta}\right|_{\theta = \theta^*}=0 ,
\end{equation}
where we call $\theta^*$ the zero-response point.
Consequently we may write
\begin{equation}
\frac{d\langle\beta W_{gen}\rangle}{d\theta} = -\int_{\theta^*}^\theta F(\theta^\prime)d\theta^\prime .
\end{equation}
In many systems the value of $\theta^*$ has a particular significance computationally, as will be demonstrated in Section~\ref{sec:thermo-analysis}.
Once $\theta^*$ is determined, we obtain:
\begin{equation}
\label{eq:theta-zero-star}
c(\theta^0) = \int_{\theta^0}^{\theta^*} F(\theta)d\theta .
\end{equation}


We demonstrate that there is another way to relate the Fisher information and the curvature of the configuration entropy.
As described in Appendix~\ref{sec:derivation-entropy-curvature}, the second derivative of the configuration entropy $S(\theta) = -\sum_x p(x|\theta)\ln p(x|\theta)$ over $\theta$, can be explicitly taken, leading to our third result
\begin{equation}
\label{eq:entropy-curv-qf}
\begin{aligned}
\frac{d^2(\mathbb{S})^+}{d\theta^2} &\equiv \frac{d^2 S}{d\theta^2} + F(\theta)\\
&= -\sum_x \frac{d^2 p(x|\theta)}{d\theta^2} \ln p(x|\theta) .
\end{aligned}
\end{equation}
Unlike Equation~\eqref{eq:fisher-quasi-static}, which captured the difference between two curvatures, Equation~\eqref{eq:entropy-curv-qf} captures the sum of two curvatures, and thus reflects a different information-geometric aspect of critical dynamics during collective motion.
Contrasting Equations~\eqref{eq:fisher-quasi-static} and ~\eqref{eq:entropy-curv-qf}, the second derivative of $\langle\beta U_{gen}\rangle$ with respect to $\theta$ can be expressed in terms of $F(\theta)$ and $(\mathbb{S})^+$ as
\begin{equation}
\label{eq:sens-flux}
\frac{d^2\langle\beta U_{gen}\rangle}{d\theta^2} = \frac{d^2(\mathbb{S})^+}{d\theta^2} - 2F(\theta_m) .
\end{equation}
In our computational analysis, which are presented in Section~\ref{sec:thermo-analysis}, we will use Equation~\eqref{eq:fisher-quasi-static}, while also showing the profile of the aggregated curvature in Equation~\eqref{eq:entropy-curv-qf}.


Finally, we propose a measure for the \emph{thermodynamic efficiency of computation}, defined as the reduction in uncertainty (i.e., the increase in order) from an expenditure of work for a given value of the control parameter:
\begin{equation}
\label{eq:ratio-comp}
\eta \equiv \frac{-dS / d\theta }{d\langle\beta W_{gen}\rangle / d\theta} = \frac{-dS / d\theta }{\int_{\theta}^{\theta^*} F(\theta^\prime)d\theta^\prime} ,
\end{equation}
which can be considered entirely in computational terms as the ratio of increasing order at $\theta$ to the cumulative sensitivity incurred over a process from $\theta$ to the zero-response point $\theta^*$.


\subsection{Simulations and probability distribution of the relative particle velocity}

\begin{figure}[b!]
\centering
\subfigure[]{\label{fig:swarm-np}\includegraphics[width=0.49\columnwidth]{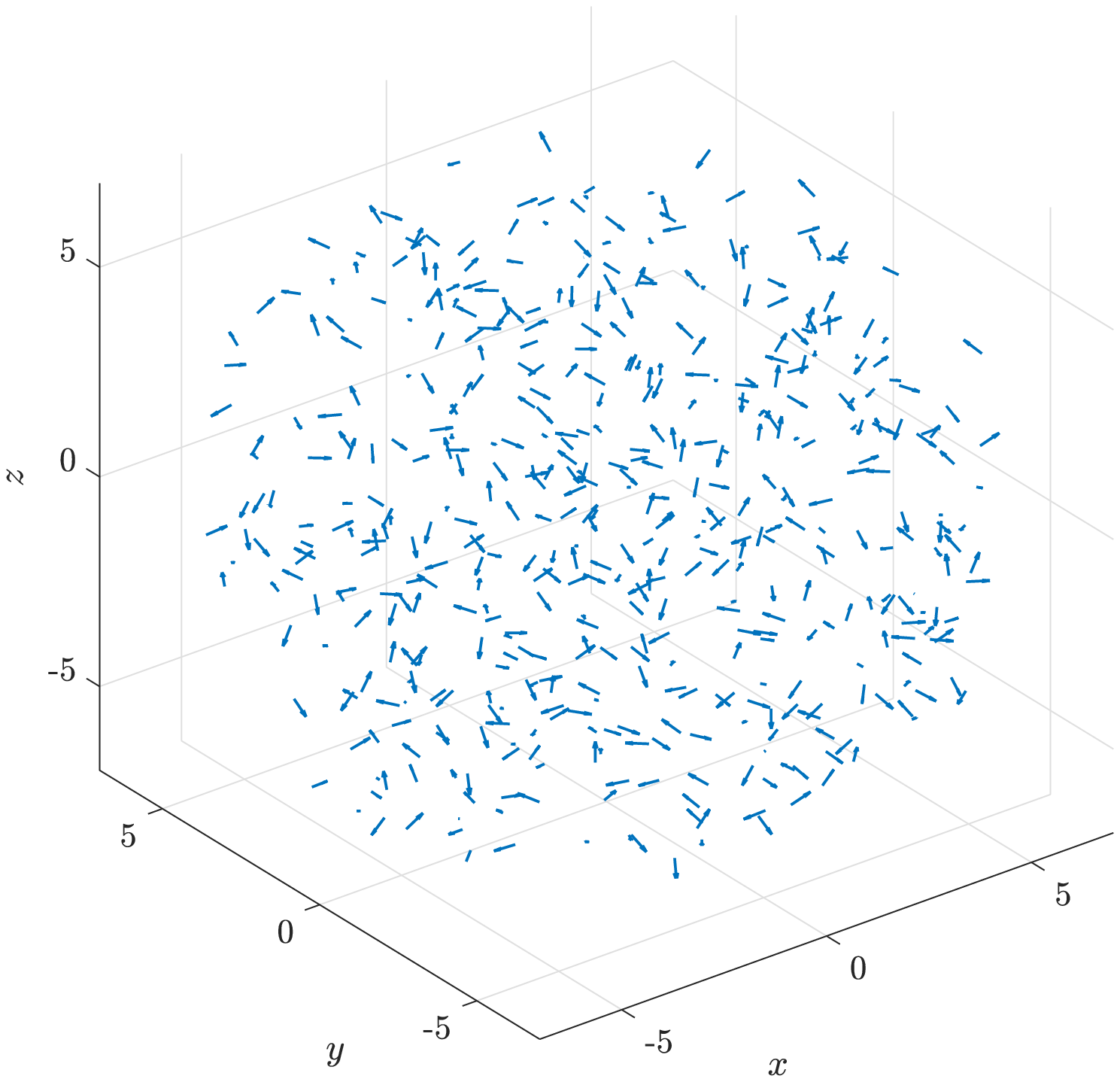}}
\subfigure[]{\label{fig:swarm-p}\includegraphics[width=0.49\columnwidth]{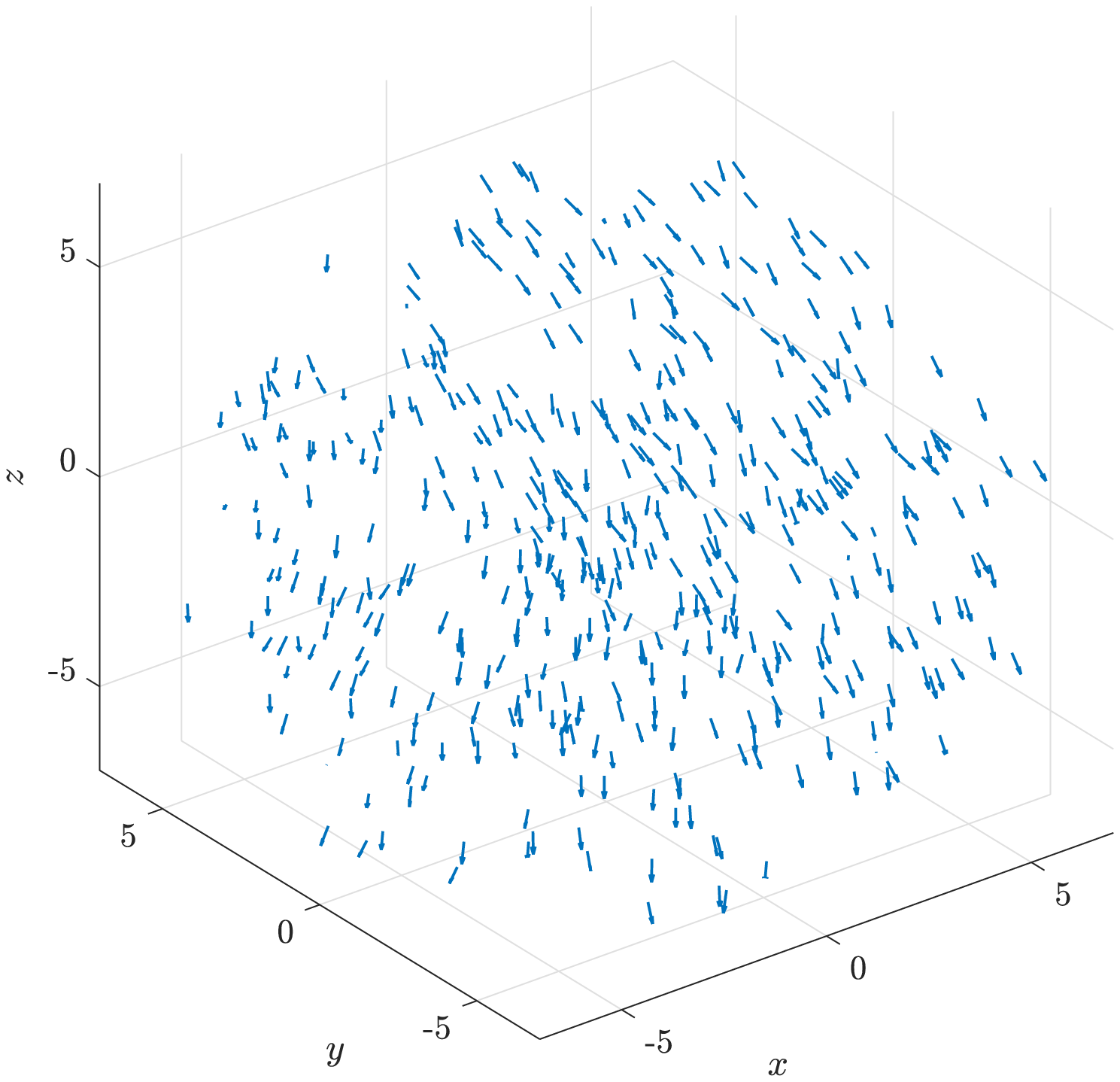}}
\caption{
Two kinetic phases of the model of collective motion.
Each arrow represents a particle with its position and velocity in the $(x,y,z)$ space.
Figure~\ref{fig:swarm-np} is taken from a simulation of the model in which $J=0.001$ and $n_c=20$, after the relaxation time, and shows the system in its disordered motion phase.
Figure~\ref{fig:swarm-p} is taken from a simulation in which $J=0.2$ and $n_c=20$, after the relaxation time, and shows the system in its coherent motion phase.
}
\label{fig:swarm}
\end{figure}


Computing the Fisher information and the entropy of a system requires the knowledge of the probability distribution $p(x|\theta)$ of the random variable, given the control parameters.
For collective motion of simulated self-propelled particles, the control parameters that we consider are the alignment strength $J$ between particles and the number of nearest neighbours $n_c$ of each particle, while the random variable that we consider is the particles' velocity $\mathbf{v}_i$ with respect to the group (assuming that the probability distribution is the same for each particle in the group).
Since in this study we consider a model of collective motion, the probability distribution of the of particles' velocity can be estimated from the simulation of the system.
Alternatively, one can, for example, follow Bialek et al.~\cite{bialek2012statistical} and estimate $p(\mathbf{v}_i|J,n_c)$ from experimental data using Equation~\eqref{eq:cm-model-bialek}.


We simulated the dynamical model~\cite{gregoire2004onset} in Equations~\eqref{eq:cm-model-position} and~\eqref{eq:cm-model-velocity} setting the weight of the alignment component to $a=Jn_c/v_0$, for several different combinations of the parameters $J$ and $n_c$, with $J$ ranging between $0.001$ and $0.2$ and $n_c$ ranging between $1$ and $30$.
In every simulation, we used $N=512$ particles and the following values of the parameters: $r_b=0.2$, $r_e=0.5$, $r_a=0.8$, $r_0=1$, $b=5$ and $v_0=0.05$.
The same setup of the model was used by Bialek et al.~\cite{bialek2012statistical} to validate their statistical mechanical model, and corresponds to the liquid phase identified by Gr\'{e}goire and Chat\'{e}~\cite{gregoire2004onset}.
We performed $100$ runs for each combination of $J$ and $n_c$ that we considered.
At the beginning of each run, the positions of the particles were randomly set within a sphere of radius proportional to the cube root of the number of particles.
The initial velocity of the particles was also randomly chosen.
During each run, the three-dimensional velocities $\mathbf{v}_i$ of each particles $i$ were recorded for $100$ time steps, after a relaxation time of $50$ that allows the system to reach the stationary state.

\begin{figure}[t]
\centering
\includegraphics[width=0.49\columnwidth]{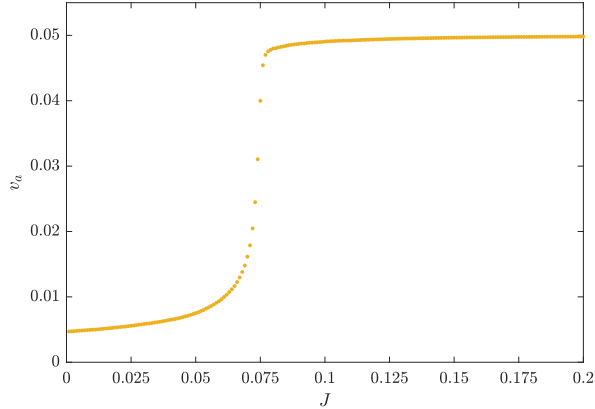}
\caption{
Average normalised velocity of the group over the alignment strength $J$.
The horizontal axis represents $J$ from $0$ and $0.2$, at steps of $0.001$, and the vertical axis represents the average normalised velocity of the group $v_a = \frac{1}{Nv_0}\left | \sum_{i=1}^N \mathbf{v}_i \right |$ over the simulation time.
The parameter $n_c$ is fixed at $20$.
}
\label{fig:phase-transition-speed}
\end{figure}

\begin{figure}[t]
\subfigure[]{\label{fig:polar-coor}\includegraphics[height=4.0cm]{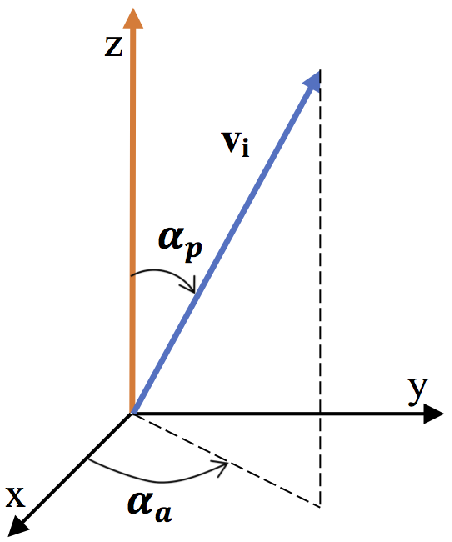}}
\subfigure[]{\label{fig:ps1}\includegraphics[height=4.2cm]{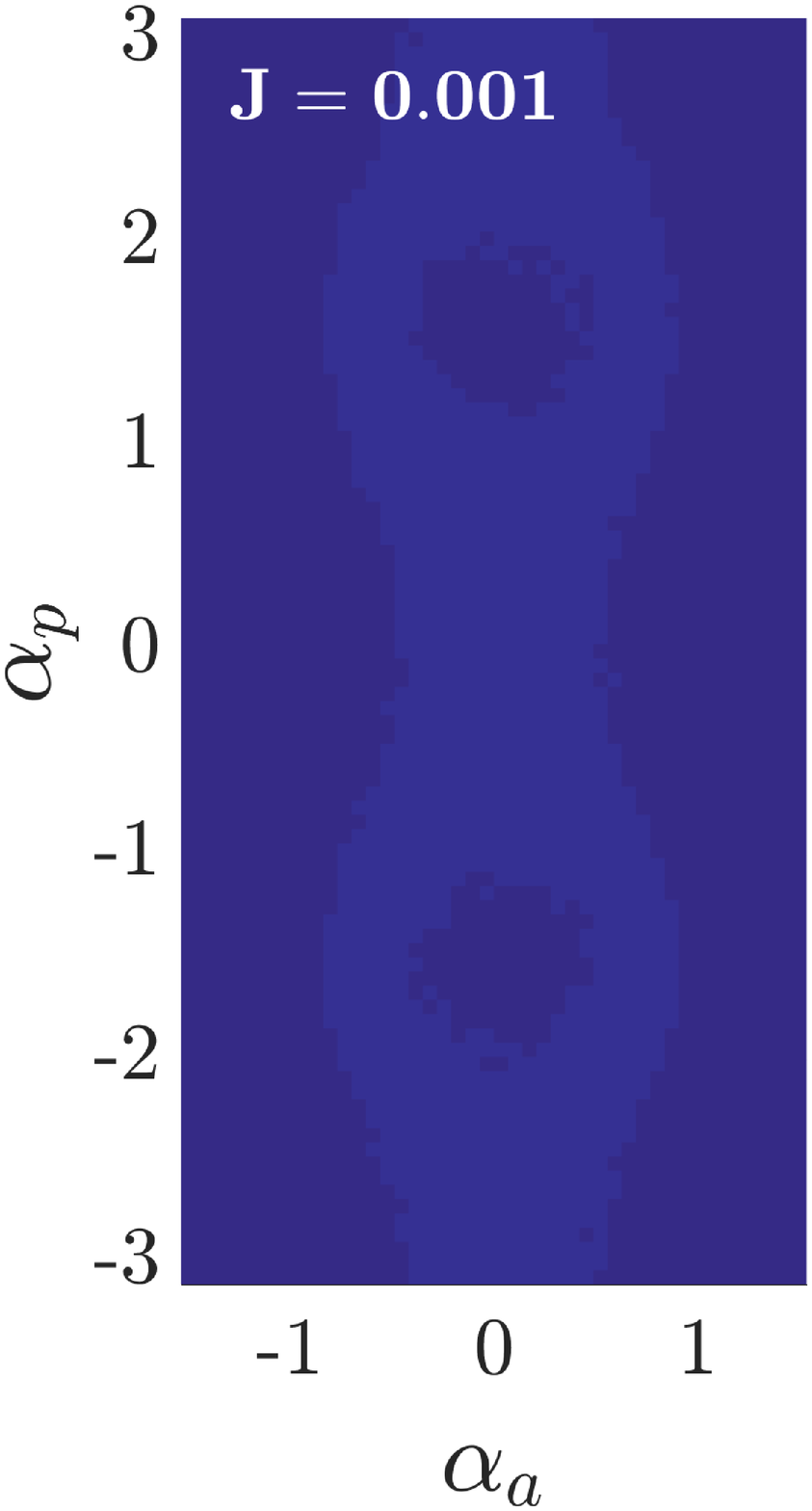}}
\subfigure[]{\label{fig:ps2}\includegraphics[height=4.2cm]{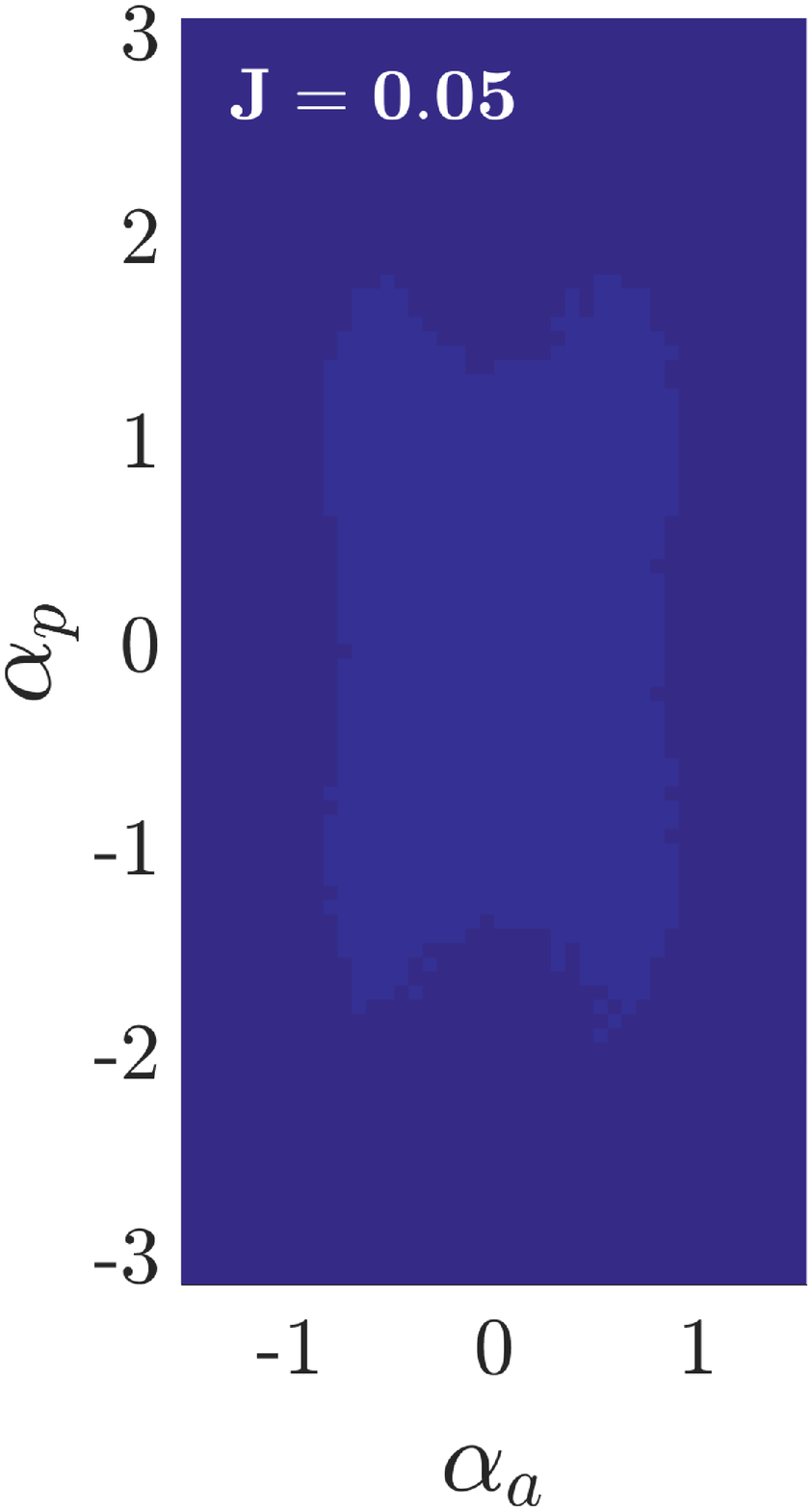}}
\subfigure[]{\label{fig:ps3}\includegraphics[height=4.2cm]{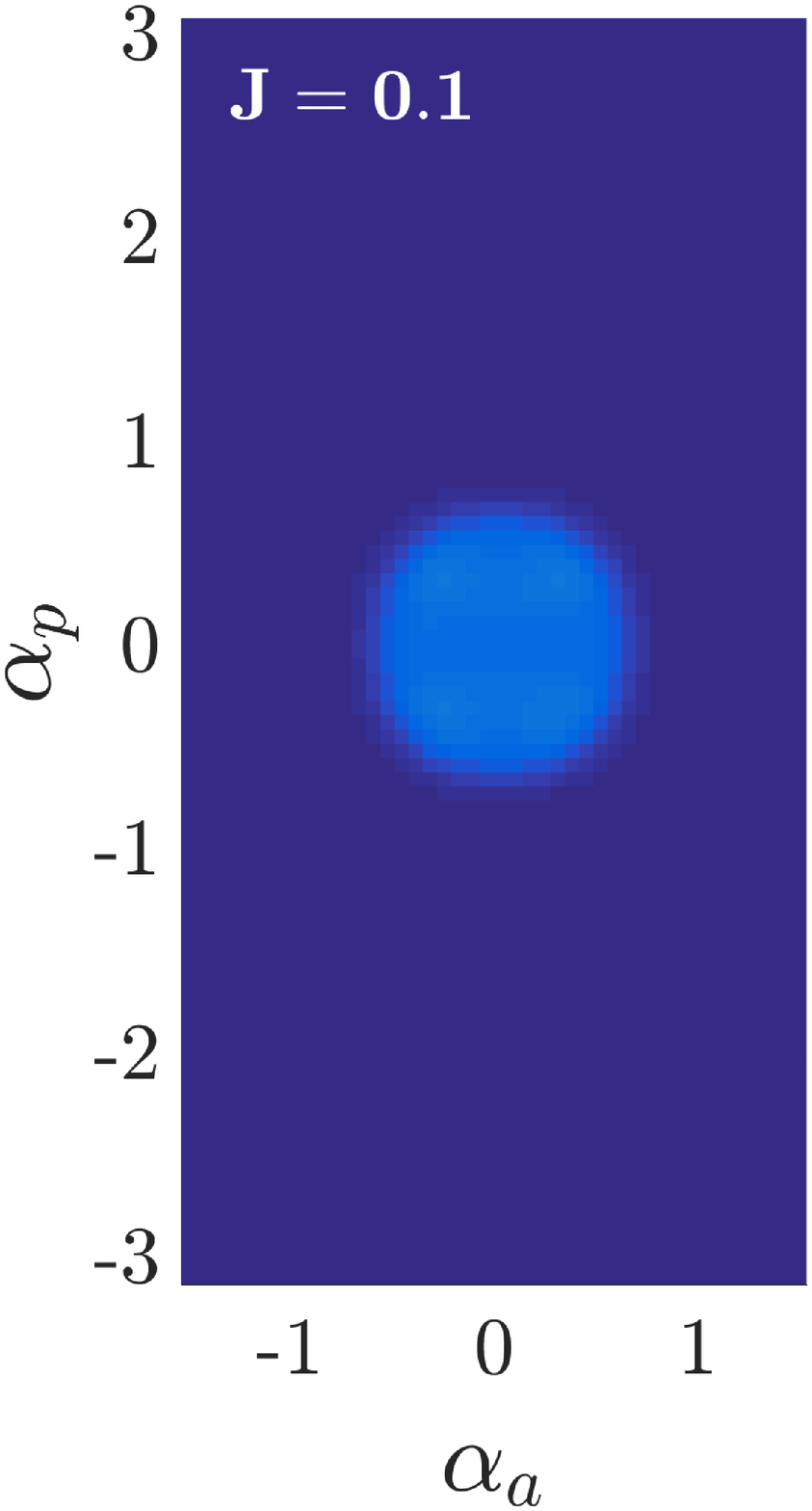}}
\subfigure[]{\label{fig:ps4}\includegraphics[height=4.2cm]{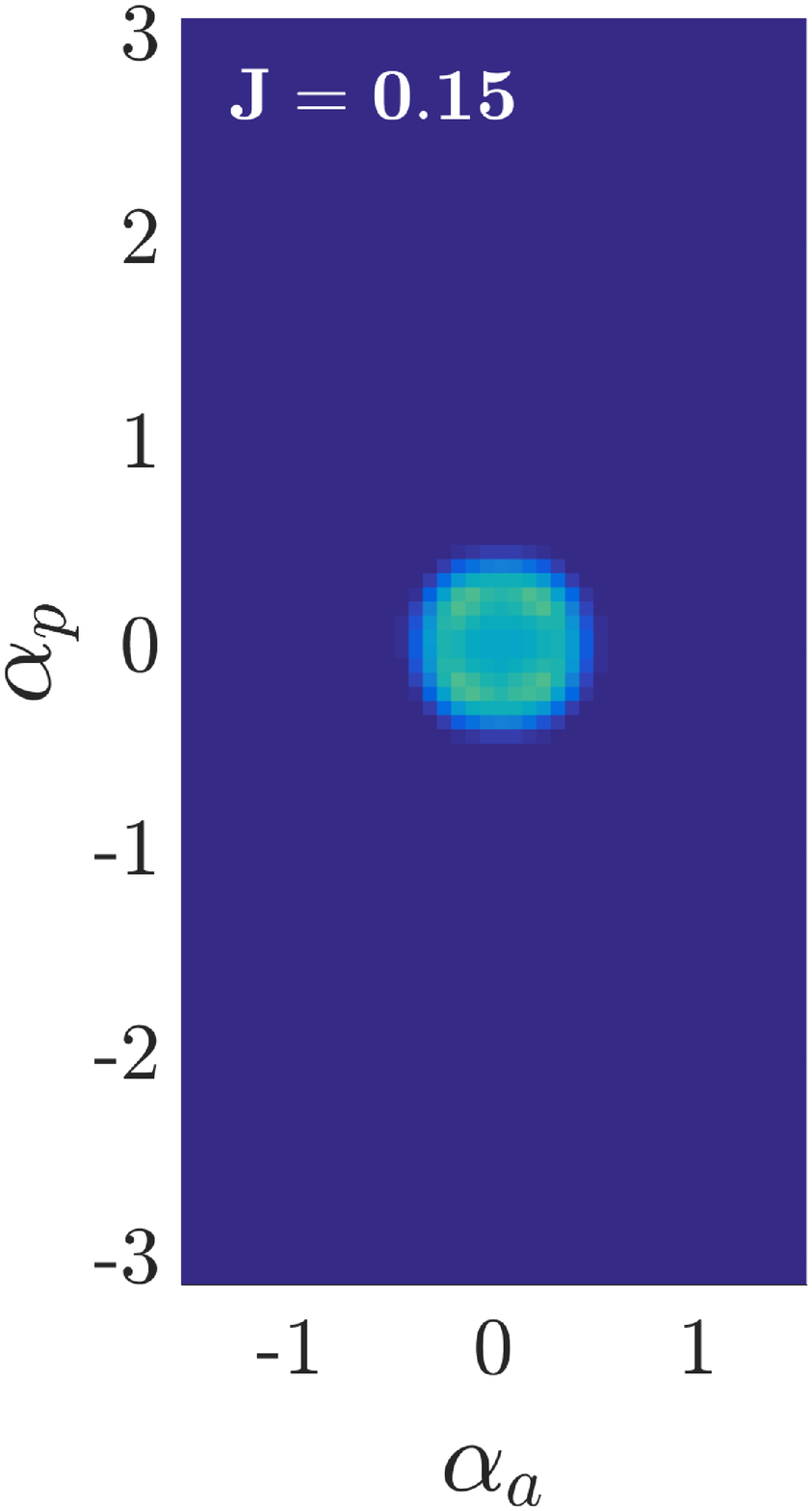}}
\subfigure[]{\label{fig:ps5}\includegraphics[height=4.45cm]{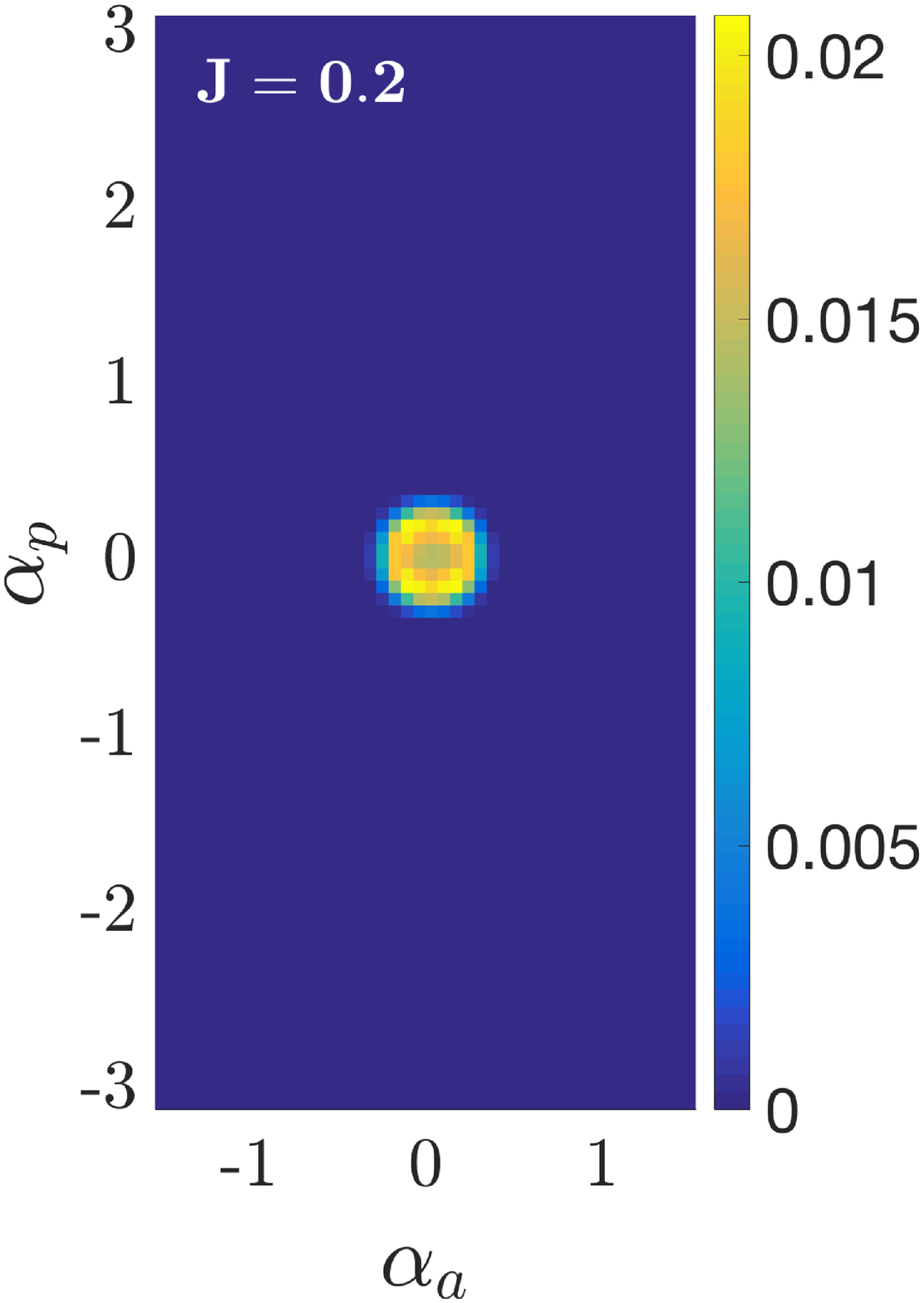}}
\caption{
Probability distribution of particles' velocity for different values of $J$.
Figure~\ref{fig:polar-coor} illustrates how a particle's velocity, with respect to the average velocity of the particles in its neighbourhood, is defined by two spherical coordinates.
The orange arrow represents the average velocity of the neighbouring particles.
A coordinate system is created so that the average velocity of the neighbourhood is the z-axis.
The vector $\mathbf{v}_i$ (blue arrow) is then the velocity of a particle with respect to this coordinate system, which can be expressed by the spherical coordinates $(\rho,\alpha_p,\alpha_a)$, where $\rho$ is the radial distance, $\alpha_p \in [-\pi,\pi]$ is the polar angle and $\alpha_a \in [-\pi/2,\pi/2]$ is the azimuthal angle.
Since in the model that we consider the speed of the particles is constant, $\alpha_p$ and $\alpha_a$ are sufficient for specifying $\mathbf{v}_i$.
Figures~\ref{fig:ps1} to~\ref{fig:ps5} show the probability distribution of the discretised cluster velocity $p(\mathbf{s}_k|J,n_c)$, for increasing values of $J$ from $0.001$ to $0.2$, and with $n_c$ fixed at $20$.
The horizontal and vertical axis are used for representing $\mathbf{s}_k$ and indicate the azimuthal and polar angles respectively, while $p(\mathbf{s}_k|J,n_c)$ is represented using a colour scale, which varies from dark blue for the lowest values, to light yellow for highest values.
}
\label{fig:ps}
\end{figure}


Running the system over a range of values of the control parameters, using relatively small changes and allowing for a relaxation time, not only enabled us to explore the behaviour of the system across the space of the control parameters, but also provided an approximation of a quasi-static protocol.
For example, all runs with the same value of $n_c$ and $J$ varying from an initial to a final value can be considered, altogether, as a quasi-static process in which a single control parameter, $J$, is varied infinitesimally slowly over time.
This approximation allows us to carry out the thermodynamical analysis described in Sections~\ref{sec:thermo-analysis}.


The simulations (see Supplemental Video 1 for a demonstration of the dynamics of the system) show that the model has two different kinetic phases of collective motion, as Gr\'{e}goire and Chat\'{e} had previously pointed out in their study~\cite{gregoire2004onset}.
In the disordered motion phase, illustrated in Figure~\ref{fig:swarm-np}, particles keep changing direction, but maintain a fairly stable collective position.
This phase corresponds to lower values of the alignment weight $a$: in the figure, for example, the parameter $J$, which is directly proportional to the alignment weight, is set to a low value of $0.001$, while $n_c$ is set to $20$.
In the coherent motion phase, illustrated in Figure~\ref{fig:swarm-p}, particles face a common general direction, and collectively move along it.
This phase corresponds to higher values of $a$: in the figure, the parameter $J$ is increased to $0.2$, while $n_c$ is again set to $20$.
The case in which $n_c$ is fixed at $20$ and $J$ varies from $0.001$ to $0.2$ is used as the main example here and throughout the rest of the article.


In order to localise the phase transition, Gr\'{e}goire and Chat\'{e} ~\cite{gregoire2004onset} (as well as Vicsek et al~\cite{vicsek1995novel}, on a previous model) utilised the order parameter
\begin{equation}
v_a = \frac{1}{Nv_0}\left | \sum_{i=1}^N \mathbf{v}_i \right | ,
\end{equation}
i.e., the absolute value of the average normalised velocity.
We inspected $v_a$ in our simulations, for different combinations of the control parameters $J$ and $n_c$.
An example is given in Figure~\ref{fig:phase-transition-speed}, which shows the average $v_a$ computed over all simulations, for each value of $J$ from $0.001$ to $0.2$ and using a fixed value of $n_c=20$.
The figure clearly reveals the phase transition: the average normalised velocity grows with the alignment strength, and the increment is particularly steep near a critical point, at approximately $J=0.075$.
A similar behaviour is observed when the alignment strength $J$ is fixed and we vary the number of nearest neighbours $n_c$.


The probability distribution of the particles' velocity was then estimated from the data collected from the simulations.
A possible choice of the random variable is the velocity of particles with respect to the average velocity over all particles, as it changes over time.
However, the average velocity over all particles is not a suitable reference for large systems (512 in our case) in the general liquid phase under consideration.
In fact, even when the group is moving coherently, subgroups of particles which are far from each other can, at least temporarily, be oriented towards different directions.
A more suitable choice of the random variable, and the one we made in this study, is the velocity of a particle with respect to the average velocity of other particles within a certain neighbourhood (such neighbourhood should not be confused with the $n_c$ nearest neighbours).
All the results presented in this paper utilise this choice of the random variable.


In order for their probability distribution to be numerically estimated, the velocities $\mathbf{v}_i$ need to be discretised.
This was done by discretising the polar and azimuthal angles $\alpha_p$ and $\alpha_a$ (see Figure~\ref{fig:polar-coor}) of the velocity into bins measuring $4\degree$ each.
For each combination of $J$ and $n_c$, we estimated the probabilities $p(\mathbf{s}_k|J,n_c)$ of $\mathbf{v}_i$ being within the cluster $\mathbf{s}_k$, where $k$ enumerates the combinations of the two bins for $\alpha_p$ and $\alpha_a$.
The probabilities $p(\mathbf{s}_k|J,n_c)$ were estimated from the velocities of all the $512$ particles, collected over all the $100$ simulations in which the combination of $J$ and $n_c$ was used, by dividing the number of recorded velocities within $\mathbf{s}_k$ by the total number of recorded velocities.
An example is given in Figure~\ref{fig:ps}, which shows $p(\mathbf{s}_k|J,n_c)$ for increasing values of $J$, from $0.001$ to $0.2$, fixing $n_c$ to $20$ (see Supplemental Video 2 for the full change of $p(\mathbf{s}_k|J,n_c)$ over $J$ at steps of $0.001$).


For lower values of $J$ between $0.001$ and $0.5$ (see Figures~\ref{fig:ps1} and~\ref{fig:ps2}), which correspond to the disordered motion phase, the probability $p(\mathbf{s}_k|J,n_c)$ is distributed almost homogeneously among all velocity clusters $\mathbf{s}_k$, indicating that the particles' velocity is only very weekly correlated with the average velocity of their neighbours.
Additionally, we can observe that within this interval of $J$, the probability distribution changes slowly.
On the contrary, as $J$ increases from $0.05$ to $0.1$ (see Figures~\ref{fig:ps2} and~\ref{fig:ps3}), the probability $p(\mathbf{s}_k|J,n_c)$ intensifies around the velocity clusters $\mathbf{s}_k$ that correspond to $\alpha_p$ and $\alpha_a$ that are closer to $0$, indicating that the velocity of a particle is now more likely to be aligned with the average velocity of its neighbours.
The change here is abrupt, with the probability distribution for $J=0.1$ (Figure~\ref{fig:ps3}) becoming clearly non-uniform.
Contrasting Figure~\ref{fig:ps} with Figure~\ref{fig:phase-transition-speed} we can see that this change happens near the critical point at $0.075$.
For higher values of $J$ from $0.1$ to $0.2$ (see Figures~\ref{fig:ps3} to~\ref{fig:ps5}), which correspond to the coherent motion phase, the probability $p(\mathbf{s}_k|J,n_c)$ keeps becoming more dense around $\alpha_p$ and $\alpha_a$ that are closer to $0$, indicating that particles increasingly intensify their alignment with their neighbours.

These observations are addressed more formally in the next section, where we show that the Fisher information can quantify the sensitivity of the probability distribution to the control parameters.


\subsection{Fisher information and the phase transition}

\begin{figure}[b!]
\centering
\includegraphics[width=0.49\columnwidth]{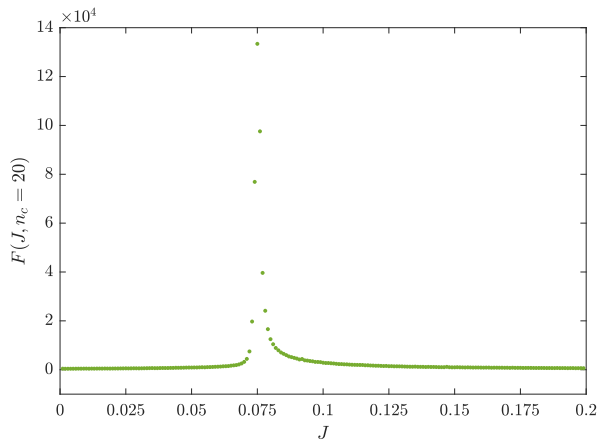}
\caption{Fisher information over the parameter $J$.
The horizontal axis represents $J$ from $0$ to $0.2$, at steps of $0.001$, and the vertical axis represents the Fisher information $F(J,n_c)$, with the parameter $n_c$ is fixed at $20$.
}
\label{fig:fisher-info}
\end{figure}

\begin{figure}[t]
\centering
\includegraphics[width=0.49\columnwidth]{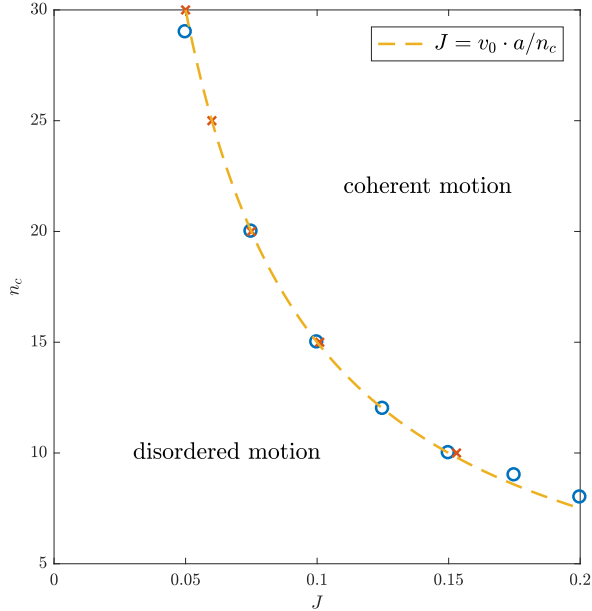}
\caption{Phase diagram using maximum Fisher information.
The horizontal axis represents the alignment strength $J$ between particles, while the vertical axis represents the number of nearest neighbours $n_c$.
Red crosses indicate the values of $J$ that yield to the higher Fisher information, for fixed values of $n_c$ from $10$ to $30$ at steps of $5$.
Analogously, blue circles indicate the values of $n_c$ that yield to the higher Fisher information, for fixed values of $J$ from $0.05$ to $0.2$ at steps of $0.25$.
The yellow dotted line is the function $J = v_0\cdot a / n_c$, with $a=30$ and $v_0=0.05$, which approximates the critical curve that separates the coherent and disordered motion phases.
}
\label{fig:phase-diagram}
\end{figure}


Fisher information allows us to quantify the amount of information that velocities carry about the control parameters $J$ and $n_c$.
Fisher information over the alignment strength $J$ can be calculated from the probabilities $p(\mathbf{s}_k|J,n_c)$ estimated from the simulations, as
\begin{equation}
\label{eq:fisher}
F(J,n_c) = \sum_{k} \frac{1}{p(\mathbf{s}_k|J,n_c)} \bigg( \frac{d\ p(\mathbf{s}_k|J,n_c)}{d J} \bigg)^2 ,
\end{equation}
having fixed the value of $n_c$.
Notice that Equation~\eqref{eq:fisher} is equivalent to Equation~\eqref{eq:fisher-matrix}, for the case in which only one control parameter is considered and the random variable is discrete.
The derivative of $p(\mathbf{s}_k|J,n_c)$ over $J$ can be computed numerically using the symmetric difference quotient two-point estimation.


We computed the Fisher information over $J$ from $0.001$ to $0.2$, at steps of $0.001$, for several fixed values of $n_c$.
In Figure~\ref{fig:fisher-info} we show the Fisher information over $J$ for our example case of $n_c=20$.
We can observe that the Fisher information is mostly low, except around the critical point of the kinetic phase transition at approximately $J=0.075$, where it diverges positively.
Analogous results were obtained using different fixed values of $n_c$.
The Fisher information was similarly computed over the number of nearest neighbours $n_c$ from $1$ to $30$, at unitary steps, for several fixed values of $J$ between $0.001$ and $0.2$.
An example is shown in Appendix~\ref{sec:over-nc}, where it is also evident that the Fisher information diverges at the critical point of the kinetic phase transition.


The divergence of the Fisher information at criticality, exemplified in a system of self-propelled particles performing collective motion, allows us to localise the critical points of the kinetic phase transition in a systematic and generic way, without relying on a specific order parameter, which may or may not be defined in general.
Thus, this method may be used to detect phase transitions in cases in which the definition of a suitable order parameter is problematic.


Having observed that the Fisher information diverges at the critical point, we can use it to create a phase diagram of the behaviour of the system, over the two control parameters $J$ and $n_c$.
Figure~\ref{fig:phase-diagram} shows the phase diagram that we obtained by finding, for several fixed values of $n_c$, the corresponding values of $J$ that yields the maximum Fisher information and, vice versa, by finding values of $n_c$ that yield the maximum Fisher information for several fixed values of $J$.
We can see that the critical combinations of $J$ and $n_c$ can be approximated by the curve $J = v_0\cdot a / n_c$ where, in this case, $a=30$.
This should not come as a surprise since, in the dynamical model used for the simulation, we set the weight of the alignment component to $a=Jn_c/v_0$.
However, the topological nature of the parameter $n_c$ makes this result non-trivial.


\subsection{Thermodynamical analysis of collective motion}
\label{sec:thermo-analysis}

\begin{figure}[t]
\centering
\includegraphics[width=0.49\columnwidth]{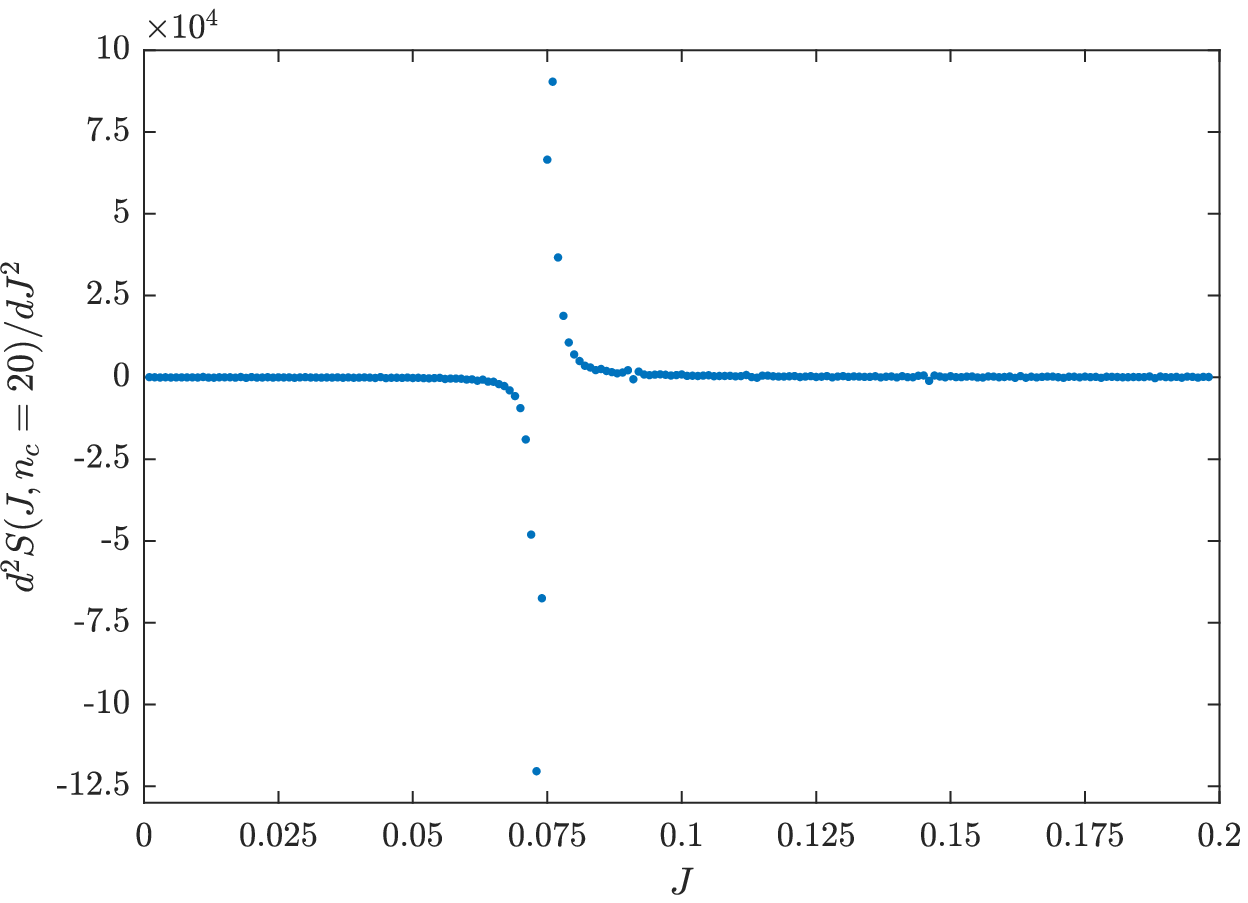}
\caption{Curvature of the configuration entropy of the system over $J$.
The horizontal axis represents $J$ from $0$ to $0.2$, at steps of $0.001$, and the vertical axis represents the curvature of the configuration entropy of the system $S(J,n_c)$, with the parameter $n_c$ is fixed at $20$.
}
\label{fig:ent-curv}
\end{figure}

\begin{figure}[t!]
\centering
\includegraphics[width=0.49\columnwidth]{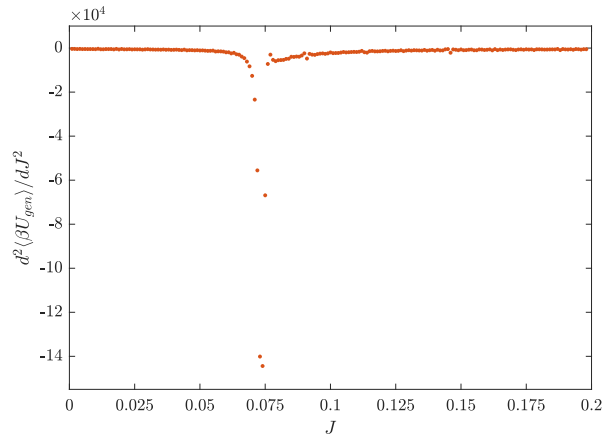}
\caption{Second derivative of the generalised internal energy with respect to $J$ ($\beta=1$).
The horizontal axis represents $J$ from $0$ to $0.2$, at steps of $0.001$, and the vertical axis represents the second derivative of the generalised internal energy with respect to $J$.
The parameter $n_c$ is fixed at $20$.
}
\label{fig:sens-entropy-flux}
\end{figure}


As described in Section~\ref{sec:fisher-and-thermo}, the Fisher information represents the negative second derivative of the generalised work done on, or extracted from, the system due to changing the control parameter in the quasi-static limit.
Therefore, Figure~\ref{fig:fisher-info} also provides, with opposite sign, the curvature of work with respect to the alignment strength $J$ (assuming $\beta=1$) for our example case in which the number of nearest neighbours $n_c$ is fixed at $20$ and $J$ varies from $0.001$ to $0.02$.
Hence, the second derivative of work diverges negatively near the critical point.

\begin{figure}[t]
\subfigure[]{\label{fig:prod-flux-conf-lines}\includegraphics[width=0.49\columnwidth]{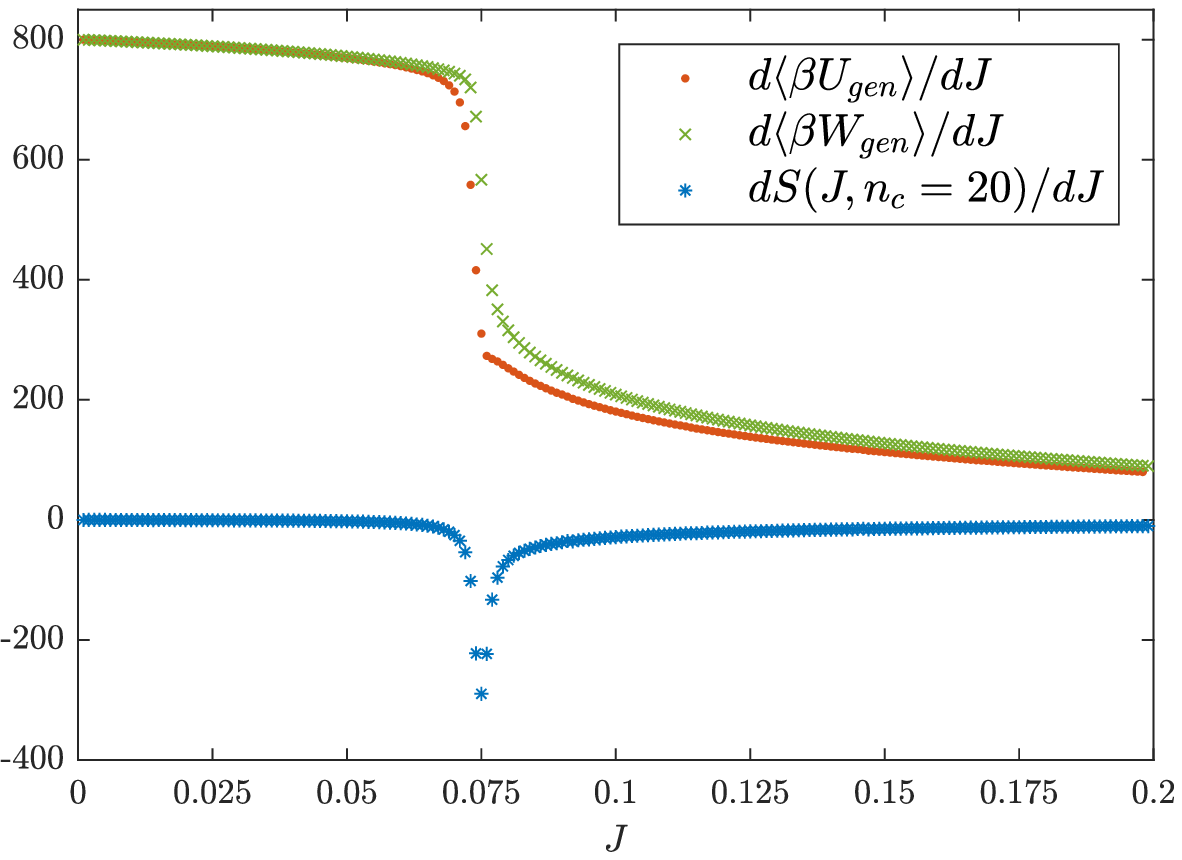}}
\subfigure[]{\label{fig:ratio}\includegraphics[width=0.49\columnwidth]{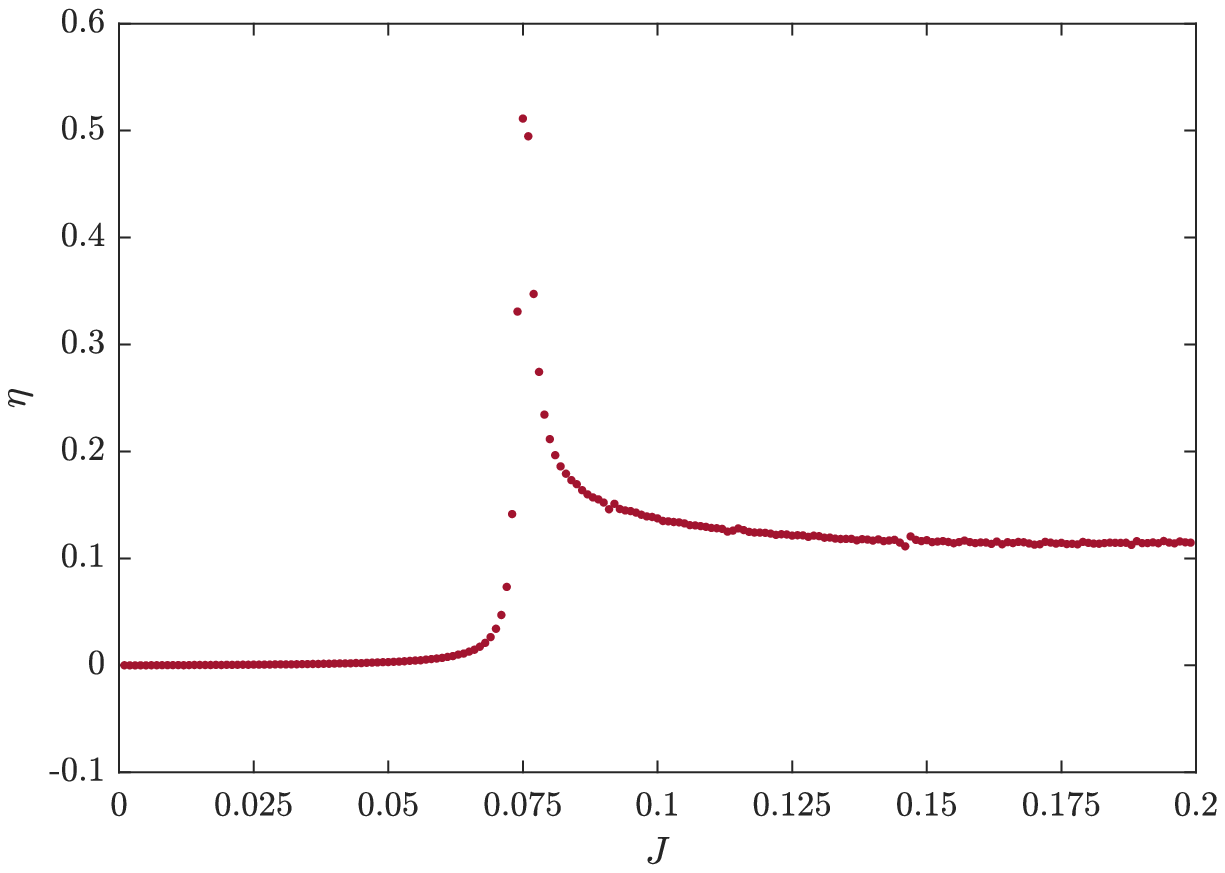}}
\caption{
Rates of change of work, internal energy and configuration entropy with respect to $J$, with $n_c$ is fixed at $20$.
In both graphs, the horizontal axis represents $J$ from $0$ to $0.2$, at steps of $0.001$.
The green crosses in Figure~\ref{fig:prod-flux-conf-lines} represent the rate of change with respect to $J$ of the generalised work ($\beta=1$), the red dots represent the rate of change with respect to $J$ of the generalised internal energy ($\beta=1$) and the blue asterisks represent the rate of change with respect to $J$ of the configuration entropy.
Figure~\ref{fig:ratio} shows the thermodynamic efficiency of computation $\eta$ over $J$.}
\label{fig:prod-flux-conf}
\end{figure}

On the other hand, the second derivative of the internal energy of the system, over a control parameter, is proportional to the difference between two curvatures: the second derivative of the configuration entropy of the system and the Fisher information (see Equation~\eqref{eq:fisher-quasi-static}).
For our system of self-propelled particles, the configuration entropy can be computed for every combination of $J$ and $n_c$ as
\begin{equation}
\label{eq:entropy-swarm}
S(J,n_c) = - \sum_{k}p(\mathbf{s}_k|J,n_c) \ln p(\mathbf{s}_k|J,n_c) .
\end{equation}
The curvature of the configuration entropy was obtained by numerically computing the second derivative of the $S(J,n_c)$ determined by Equation \eqref{eq:entropy-swarm}, over the parameter $J$, using the symmetric difference quotient two-point estimation.
The result is shown in Figure~\ref{fig:ent-curv}, while the configuration entropy itself is shown in Appendix~\ref{sec:entropy-and-derivative} and its first derivative can be seen in Figure~\ref{fig:prod-flux-conf-lines}.
It can be observed that the curvature of the configuration entropy is also mostly low, except near the critical point at $J=0.075$, where it diverges negatively from the left and positively from the right, thus exhibiting a discontinuity.


Applying Equation~\eqref{eq:fisher-quasi-static}, we can calculate the second derivative of the internal energy (scaled by $\beta$) with respect to $J$ as the difference between the Fisher information in Figure~\ref{fig:fisher-info} and the curvature of the configuration entropy in Figure~\ref{fig:ent-curv}, that is
\begin{equation}
\label{eq:flux-sens-j}
\frac{d^2\langle\beta U_{gen}\rangle}{d J^2} = \frac{d^2 S(J,n_c)}{dJ^2} - F(J, n_c) ,
\end{equation}
yielding the result in Figure~\ref{fig:sens-entropy-flux}.
It can be observed that the second derivative of $\langle\beta U_{gen}\rangle$ also diverges at the critical point of the phase transition.
In fact, it changes over $J$ similarly to the second derivative of the generalised work (the opposite of the Fisher information) in Figure~\ref{fig:fisher-info}.


If we consider the system of self-propelled particles as a system that performs distributed computation during collective motion, Figures~\ref{fig:fisher-info},~\ref{fig:ent-curv} and~\ref{fig:sens-entropy-flux} will reveal a computational balance between the sensitivity and the uncertainty of the computation.
On the one hand, the sensitivity of the system to changes in the control parameter is captured by the Fisher information in Figure~\ref{fig:fisher-info}.
On the other hand, the uncertainty of the computation is captured by the curvature of the configuration entropy of the system in Figure~\ref{fig:ent-curv}.
In either the disordered motion phase, or the coherent motion phase, Figure~\ref{fig:sens-entropy-flux} shows that there is a balance between the sensitivity and the uncertainty, but it is clear that this balance is broken at criticality.


\begin{figure}[t]
\centering
\includegraphics[width=0.49\columnwidth]{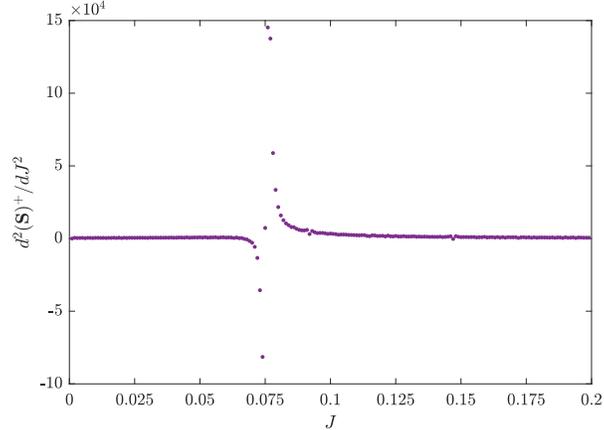}
\caption{Aggregated curvature $d^2(\mathbb{S})^+ / dJ^2$.
The horizontal axis represents $J$ from $0$ to $0.2$, at steps of $0.001$, and the vertical axis represents the aggregated curvature.
The parameter $n_c$ is fixed at $20$.
}
\label{fig:Q}
\end{figure}

The sum of the Fisher information and the curvature of the configuration entropy was also inspected.
For the system of self-propelled particles, this quantity varying over $J$ is determined as
\begin{equation}
\frac{d^2(\mathbb{S})^+}{dJ^2} = -\sum_k \frac{d^2 p(\mathbf{s}_k|J,n_c)}{dJ^2} \ln  p(\mathbf{s}_k|J,n_c)
\end{equation}
and is shown in Figure~\ref{fig:Q}.
It can be observed that this quantity also has a discontinuity at criticality, similarly to the curvature of the configuration entropy.


The rate of change over $J$ of the generalised work ($\beta=1$) and the generalised internal energy ($\beta=1$) can be obtained by numerically integrating the curvatures of these quantities over $J$.
As explained in Section~\ref{sec:fisher-and-thermo}, this can only be calculated if the integration starts from a point where the work rate, or the internal energy rate, with respect to $J$ is known.
In our case, we assert that the zero-response point $J^*$ is realised as $J\to \infty$, since in this region $dJ$ produces no work, because all the particles are already perfectly aligned.
Consequently, we associate the zero-response point with the state of perfect order.
In our case, we have $J^*=\infty$ and choose $J^0=0$, which according to Equation~\eqref{eq:theta-zero-star} yields
\begin{equation}
c(0) = \int_{0}^\infty F(J,n_c)dJ
\end{equation}
and
\begin{equation}
\frac{d\langle\beta W_{gen}\rangle}{dJ} = -\int_{0}^J F(J',n_c)dJ' + c(0) .
\end{equation}

Computing $c(0)$ requires a numerical estimation, which we approximated to have a lower bound, $c(0)>800$, for $n_c=20$.
This is reflected in all plots.
The integration was done using the cumulative trapezoidal numerical integration, and the result is shown in Figure~\ref{fig:prod-flux-conf-lines}.
We can see that the rate of change of the generalised work (green crosses) decreases with $J$.
Figure~\ref{fig:prod-flux-conf-lines} also shows the first derivative of the configuration entropy over $J$ (blue asterisks).
As we can see, the configuration entropy decreases around the critical point, where the system of self-propelled particles self-organises in a more ordered phase and begins to display coherent collective motion.

Importantly, as the alignment strength $J$ increases, the entropy decreases and the work rate is positive: generating order requires work to be expended.
Specifically at the critical point we find that the ratio of generated order to the work rate peaks, indicating that the maximal thermodynamical efficiency of computation carried out by the system of self-propelled particles, that is
\begin{equation}
\eta=\frac{-dS(J, n_c)/dJ}{d\langle\beta W_{gen}\rangle/dJ} = \frac{-dS(J, n_c) / dJ }{\int_{J}^{\infty} F(J', n_c)dJ'}
\end{equation}
is the highest at criticality (see Figure~\ref{fig:ratio}).
Explicitly, in computational terms, the maximum thermodynamic efficiency corresponds to a maximal ratio of generated order to the sensitivity accumulated over a process running from the current state to the state of perfect order (the zero-response point).
Since the Fisher information is always positive, the denominator $\int_{J}^{\infty} F(J')dJ'$ can be interpreted as a measure of distance, along the trajectories of $J$, from the perfectly ordered state.
Thus, it scales the increase in order as the control parameter changes.
For example, achieving one bit of uncertainty reduction near the state of perfect order is much more significant than achieving one bit of uncertainty reduction in a largely disordered state.
This means that at criticality, the reduction of uncertainty is the most significant, reflected in the highest thermodynamic efficiency of computation.

A similar analysis was conducted for the case in which the alignment strength $J$ is fixed and the parameter $n_c$ changes.
The results are presented in Appendix~\ref{sec:over-nc}.


\section{Discussion and conclusions}
\label{sec:conclusions}


In this study we presented a theoretical framework for measuring fundamental thermodynamical quantities, such as the generalised heat, work and energy, in systems driven by quasi-static protocols.
Importantly, such framework relates these thermodynamical quantities to information-theoretic ones, and specifically to the configuration entropy and the Fisher information.
We applied the framework to a system of simulated self-propelled particles during collective motion, and studied the (generalised) internal energy and work done on, or extracted from, the system as it goes through a kinetic phase transition.


The model of collective motion that we considered is the one proposed by Gr\'{e}goire and Chat\'{e}~\cite{gregoire2004onset}, which is known to have a kinetic phase transition over control parameters influencing the particles' alignment: from a disordered motion phase, in which particles maintain a fairly stable collective position, to a coherent motion phase, in which particles cohesively move towards a common direction.
We have considered two control parameters, i.e., the alignment strength between particles and the number of nearest neighbours influencing the particles' alignment, within intervals in which the kinetic phase transition is observed.
In order to approximate a quasi-static protocol, we simulated the system for chosen values of the control parameters, under the assumption that the system reaches a stationary state after a certain relaxation time, and we repeated the experiments for different values of the control parameters.
We also used the data collected with the simulations to numerically estimate the probability distribution of the velocity of the particles at different values of the control parameters.


Our approach involves a statistical mechanical formulation of the second derivatives of the generalised internal energy and the generalised work with respect to the control parameters, based on relationships between these quantities and two other quantities, the Fisher information and the curvature of the configuration entropy, which can be calculated from the probability distribution of the velocities.
Additionally, our method provides an information-geometric interpretation of the curvature of the internal energy of the system (scaled by $\beta$) as the difference between two curvatures: the curvature of the free entropy, captured by the Fisher information, and the curvature of the configuration entropy (Equation~\eqref{eq:fisher-quasi-static}).
Another expression (Equation~\eqref{eq:entropy-curv-qf}), also interpreted information-geometrically as an aggregated curvature, is given for the sum of the Fisher information and the curvature of the configuration entropy of the system.


The expression representing the difference between curvatures (Equation~\eqref{eq:fisher-quasi-static}) highlights the computational balance between the sensitivity of the computation, captured by the Fisher information, and the uncertainty of the computation, captured by the configuration entropy, that is performed by the system of self-propelled particles during collective motion.
Our numerical results show that such balance is stressed at criticality, where the curvatures with respect of the control parameters of the generalised work and the generalised internal energy, as well as the curvature of the configuration entropy of the system, diverge.
The rates of change of the generalised work and the generalised internal energy over the control parameters were estimated from the curvatures of these quantities, using numerical integration.
The results show that during the kinetic phase transition, when the configuration entropy of the system decreases very rapidly, both the 
rate of change of the work and the internal energy decrease dramatically.


Our results support the view that flocking behaviour, which combines coherence and responsiveness to external perturbations (e.g., predatory attacks), exhibits criticality in the statistical mechanical sense~\cite{cavagna2010scale, mora2011biological, bialek2012statistical, bialek2014social}.
Moreover, our results suggest that the highest thermodynamic efficiency of computation $\eta$, relating the reduction of the configuration entropy to the required work rate, is achieved at criticality.
We have also shown that this quantity can be interpreted as a significance of entropy reduction with respect to the distance along a computational path to a perfectly ordered state, where the distance is understood to mean the cumulative sensitivity captured by the integral of the Fisher information.


When applying the proposed theoretical framework, it is crucial to imbue the derivative with respect to the control parameter with physical meaning.
In this study, we have considered the most natural case of a quasi-static protocol, however, less conventional alternatives can be constructed.
For instance, one can think of a feedback process, in which the protocol is changed in response to measurements of the process~\cite{toyabe2010experimental, sagawa2010generalized, sagawa2012nonequilibrium}.
 If the measurements gain equal or more information than the free energy change, then the measurement can be used to change the protocol so that zero work is performed (or extracted) upon changing the control parameter.
 Because of the first law of thermodynamics, if no work is done then $\Delta\langle U_{gen}\rangle = \Delta\langle Q_{gen}\rangle$ which, following Equation~\eqref{eq:flux-sens-j} or~\eqref{eq:fisher-quasi-static}, leads to:
\begin{equation}
\label{eq-difference-heat}
\begin{aligned}
\frac{d^2(\mathbb{S})^-}{dJ^2} &= \frac{d^2\langle\beta Q_{gen}\rangle}{dJ^2}\\
&=\frac{d^2 S(J, n_c)}{dJ^2} - F(J, n_c) .
\end{aligned}
\end{equation}
Thus, the thermodynamic interpretation of $d^2(\mathbb{S})^- / dJ^2$, provided by Equation~\eqref{eq:fisher-quasi-static}, changes: it is no longer the curvature of the generalised internal energy of the system (scaled by $\beta$).
It is instead the curvature of the heat (scaled by $\beta$), which can be interpreted as the rate of change of the entropy flux $\Phi_J$ from the system to the environment~\cite{niven2010minimization, prokopenko2015information}:
\begin{equation}
\Phi_J = \int\frac{d^2\langle\beta Q_{gen}\rangle}{dJ^2}dJ .
\end{equation}

If we assume that the whole system, which includes the self-propelled particles as well as the environment, is isolated, then its total entropy production $\Pi_J$ is the difference between the rate of change of the configuration entropy of the system of self-propelled particles and the entropy flux to the environment (given the sign convention):
\begin{equation}
\Pi_J = \frac{dS(J,n_c)}{dJ} - \Phi_J .
\end{equation}
In light of this relationship, integrating Equation~\eqref{eq-difference-heat} leads to the interpretation of the Fisher information as the rate of change of the total entropy production with respect to the control parameter:
\begin{equation}
\Pi_J = \int F(J,n_c)d(J).
\end{equation}

Hence, if we look again at Figure~\ref{fig:prod-flux-conf-lines}, but this time considering the feedback process, it is clear that increasing $J$ would lead to a negative spike in total entropy production because the information has been used to reduce the work done on the system, thus decreasing irreversibility.
In contrast, for decreasing $J$ a positive spike would be observed because the information is being used counterproductively to extract zero work when positive work could be extracted, increasing irreversibility.
Interestingly, the ratio of the rate of change of the configuration entropy of the system to the total entropy production (see Figure~\ref{fig:ratio}) would be highest at criticality.

Total entropy production and entropy flux have been studied in a variety of systems, including the majority-vote model~\cite{crochik2005entropy}, copolymerisation processes~\cite{andrieux2008nonequilibrium}, a population model~\cite{andrae2010entropy}, interacting lattice gas~\cite{tome2010entropy} and the Ising model~\cite{oliveira2011irreversible, zhang2016critical}, among others.
All these studies have identified phase transitions over some control parameter (for instance, the temperature and the coupling constant were chosen as control parameters in the Ising model~\cite{oliveira2011irreversible, zhang2016critical}).
The theoretical framework proposed in this study could be applied to a range of processes in which it can be assumed that no work is done on, or extracted from, the system.


In addition to our main results, we have also shown that the critical points of the kinetic phase transition are captured by the divergence of the Fisher information.
This allowed us to use this measure to construct a phase diagram of the dynamics of the system for different combinations of the two control parameters considered, showing the critical regime separating the coherent and disordered motion phases.


Broadly, our results contribute to ``information thermodynamics'', an emerging field exploring relationships between information processing and its thermodynamic costs~\cite{esposito2011second, deffner2013information, barato2014efficiency, horowitz2014second, horowitz2014thermodynamics, parrondo2015thermodynamics, prokopenko2015fisher, spinney2016transfer, spinney2017transfer}.
These relationships are of particular interest for complex systems which need to perform their distributed computation efficiently~\cite{lizier2012coherent, prokopenko2013thermodynamic, prokopenko2014transfer, kolchinsky2017dependence, kempes2017thermodynamic}.
We hope that our work would contribute towards a unified theory of collective motion drawing on statistical mechanics and information thermodynamics, applicable to diverse collective motion phenomena including active matter~\cite{popkin2016physics, dileonardo2016active}.

\section*{Acknowledgements}
E.C. was supported by the University of Sydney's ``Postgraduate Scholarship in the field of Complex Systems'' from Faculty of Engineering \& IT and by a CSIRO top-up scholarship.
J.L. was supported through the Australian Research Council DECRA grant DE160100630.
All authors were supported by The University of Sydney's DVC Research Strategic Research Excellence Initiative (SREI-2020) project, ``CRISIS: Crisis Response in Interdependent Social-Infrastructure Systems'' (IRMA 194163).
The authors acknowledge the University of Sydney HPC service at The University of Sydney for providing HPC resources that have contributed to the research results reported within this paper.

\appendix


\section{Derivation of the curvature of the system's entropy}
\label{sec:derivation-entropy-curvature}

The Fisher information $F(\theta)$ can be related to the second derivative of the configuration entropy $S(\theta)$ as follows.
The first derivative of $S(\theta)$ over $\theta$ is
\begin{equation}
\begin{aligned}
\frac{d  S}{d \theta} &= -\sum_x \frac{d  (p(x|\theta)\ln p(x|\theta))}{d \theta}\\
&=  -\sum_x \bigg( \frac{d\ p(x|\theta)}{d \theta}\ln p(x|\theta) + \frac{d\ p(x|\theta)}{d \theta} \bigg)\\
&= -\sum_x \frac{d\ p(x|\theta)}{d \theta}\ln p(x|\theta) -\sum_x \frac{d\ p(x|\theta)}{d \theta}\\
&= -\sum_x \frac{d\ p(x|\theta)}{d \theta}\ln p(x|\theta) -\frac{d \sum_x  p(x|\theta)}{d \theta}\\
&= -\sum_x \frac{d\ p(x|\theta)}{d \theta}\ln p(x|\theta) .
\end{aligned}
\end{equation}
The second derivative $S(\theta)$ over $\theta$ is
\begin{equation}
\label{eq:second-der-ent}
\begin{split}
\frac{d ^2 S}{d \theta^2} &= -\sum_x \frac{d \Big( \frac{d\ p(x|\theta)}{d \theta} \ln p(x|\theta)\Big)} {d \theta}\\
&= -\sum_x \frac{d ^2\ p(x|\theta)}{d \theta^2} \ln p(x|\theta)\\&\ \ \ -\sum_x \frac{1}{p(x|\theta)} \bigg(\frac{d\ p(x|\theta)}{d \theta}\bigg)^2\\
&= -\sum_x \frac{d ^2\ p(x|\theta)}{d \theta^2} \ln p(x|\theta) - F(\theta).
\end{split}
\end{equation}
Comparing Equation~\eqref{eq:second-der-ent} with the definition of $\frac{d^2(\mathbb{S})^+}{d\theta^2}$ given by equation~\eqref{eq:entropy-curv-qf}, that is,
\begin{equation}
\frac{d^2(\mathbb{S})^+}{d\theta^2} \equiv \frac{d^2 S}{d\theta^2} + F(\theta)
\end{equation}
yields:
\begin{equation}
\frac{d^2(\mathbb{S})^+}{d\theta^2} = -\sum_x \frac{d^2 p(x|\theta)}{d\theta^2} \ln p(x|\theta) .
\end{equation}


\section{Configuration entropy over the alignment strength}
\label{sec:entropy-and-derivative}

\begin{figure}[t]
\centering
\includegraphics[width=0.49\columnwidth]{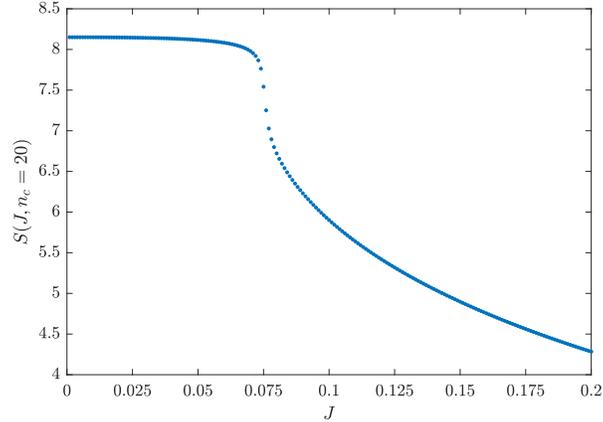}
\caption{
Entropy of the system over $J$.
The horizontal axis represents $J$ from $0$ to $0.2$, at steps of $0.001$, and the vertical axis represents the entropy of the system $S(J,n_c)$, with the parameter $n_c$ is fixed at $20$.
}
\label{fig:entropy-all}
\end{figure}

Figure~\ref{fig:entropy-all} shows that the configuration entropy of the system decreases with $J$, as the group becomes more polarised towards a flocking direction, with the drop being particularly steep in the proximity of the critical point.

\section{Entropy production and flux over the number of nearest neighbours}
\label{sec:over-nc}

\begin{figure}[h!]
\centering
\includegraphics[width=0.49\columnwidth]{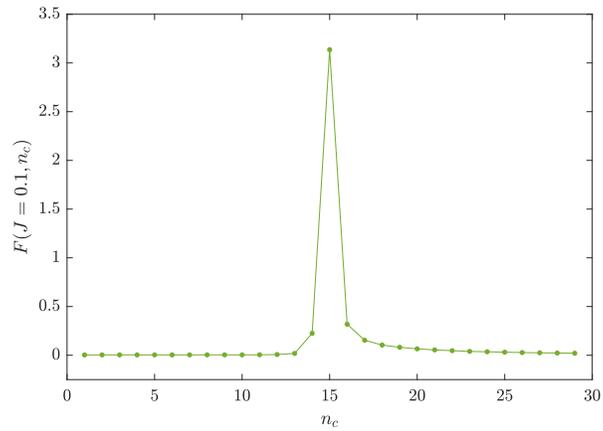}
\caption{
Fisher information over $n_c$.
The horizontal axis represents $n_c$ from $1$ to $30$, and the vertical axis represents the Fisher information $F_{n_c}(J,n_c)$, with the parameter $J$ is fixed at $0.1$.
}
\label{fig:fisher-info-nc}
\end{figure}

\begin{figure}[!]
\centering
\includegraphics[width=0.49\columnwidth]{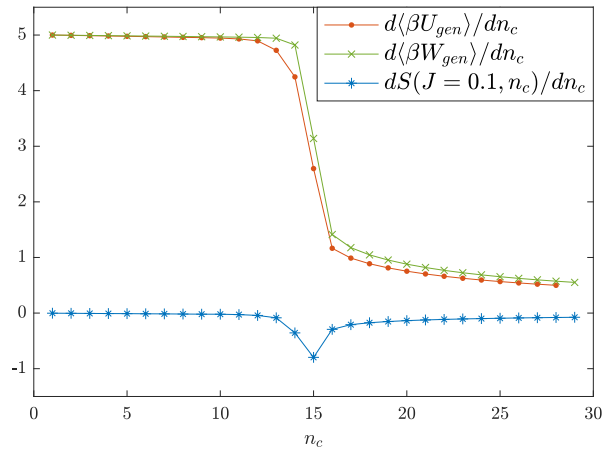}
\caption{
Rates of change of work, internal energy and configuration entropy with respect to $n_c$, with $J$ is fixed at $0.1$.
The horizontal axis represents $n_c$ from $1$ to $30$.
The green crosses represent the rate of change with respect to $n_c$ of the generalised work ($\beta=1$), the red dots represent the rate of change with respect to $n_c$ of the generalised internal energy ($\beta=1$) and the blue asterisks represent the rate of change with respect to $n_c$ of the configuration entropy.
}
\label{fig:prod-flux-conf-nc}
\end{figure}

The thermodynamical analysis in Section~\ref{sec:thermo-analysis} has also been carried out changing the control parameter $n_c$ (the number of nearest neighbours affecting the alignment component of particles motion), while fixing the control parameter $J$ (the alignment strength).
Analogous results to varying $J$ while fixing $n_c$ have been obtained, some of which are shown in this appendix.
Figure~\ref{fig:fisher-info-nc} shows that the Fisher information, which also represents the opposite of the curvature of the generalised work (scaled by $\beta$) with respect to the number of neighbours, diverges at the critical point $n_c=15$.
Figure~\ref{fig:prod-flux-conf-nc} shows the rates of change with respect to the number of nearest neighbours of the generalised work ($\beta=1$), the generalised internal energy ($\beta=1$) and the configuration entropy.
Computing $c(n_c=1)$ requires a numerical estimation, which we approximated to have a lower bound $c(1)>5$ for $J=0.1$.
The rates of change of work and internal energy decrease with $n_c$, and the drop is particularly steep at criticality.
The rate of change of the configuration entropy is instead generally low, apart from near the critical point, where it drops.

\newpage
\bibliographystyle{unsrt}


\end{document}